\documentclass[fleqn,10pt]{wlscirep}
\usepackage[utf8]{inputenc}
\usepackage[T1]{fontenc}

\pdfoutput=1
\usepackage{graphicx}
\usepackage[utf8]{inputenc}
\usepackage[english]{babel}
\usepackage[T1]{fontenc}
\usepackage{amsmath}
\usepackage{hyperref}
\usepackage{verbatim}
\usepackage{amsthm}
\usepackage{tikz}
\usepackage{lipsum}
\usepackage{braket}
\usepackage{colortbl}
\usepackage{pstricks}
\usepackage{calc}
\usepackage{subcaption}
\usepackage{booktabs}
\usepackage{caption}
\title{Single-shot detection limits of quantum illumination with multi-qudit states}

\author[1]{Sunghwa Kang}
\author[1]{Yonggi Jo}
\author[1]{Jihwan Kim}
\author[1]{Zaeill Kim}
\author[1]{Duk Y. Kim}
\author[1,*]{Su-Yong Lee}
\affil[1]{Agency for Defense Development, Daejeon 34186, KOREA}

\affil[*]{suyong2@add.re.kr}

\begin{abstract}
    Quantum illumination is a protocol for detecting a low-reflectivity target by using two-mode entangled states composed of signal and idler modes, which can outperform unentangled states. 
    We study multi-qudit states for single-shot detection limits of quantum illumination under white noise environment.
    Using three-qubit states, we obtain that the performance is enhanced by the entanglement between signal and idler qubits, whereas it is degraded by the entanglement between signal qubits. The similar behaviors are also observed for three-qutrit, four-qubit, and four-ququart states. In particular, the optimal state is not a maximally entangled multipartite state but a combination of a maximally entangled bipartite state.
    Moreover, we show that quantum correlation can explain the quantum advantage of three-qubit, three-qutrit, and four-qubit states, with exception of a four-ququart state.
\end{abstract}
\begin{document}

\flushbottom
\maketitle
%
%
\thispagestyle{empty}

\section*{Introduction}

Qubit \cite{qubit}, which represents a two-dimensional quantum mechanical system, is a fundamental unit of quantum information. 
A qubit is generalized to a $d$-dimensional system, called a qudit.
By increasing the number of qub(d)its, it is also possible to describe entanglement, which is the most prominent quantum-mechanical phenomenon \cite{EPR}.  
Entanglement is an essential ingredient of quantum teleportation, quantum computation, and quantum sensing. 
The quantum information protocols take quantum advantage over the classical limits, since entanglement is not broken during the dynamics. 
However, quantum illumination (QI) \cite{Lloyd08} can take quantum advantage over the classical limit, even if an initially prepared entanglement is broken during the dynamics. 

QI is on the purpose of distinguishing the presence and absence of a low-reflectivity target using entangled states that consist of signal and idler modes, where the target is embedded in background noise.
At a transmitter, initially, a signal mode is sent to a target while keeping an idler mode intact. 
Then, at a receiver, a reflected signal is measured with the idler mode. By processing the measurement outcomes, we can obtain a detection error probability that is a sum of false-alarm probability and miss-detection probability. The minimum detection error probability is lower bounded by the Helstrom bound (HB) and upper bounded by quantum Chernoff bound (QCB) \cite{Audenaert07,Calsa08,Stefano08}, where QCB presents the asymptotic decay rate of the error probability. Due to mathematical complexity, it is preferred to calculate QCB rather than HB. In terms of the QCB, Lloyd \cite{Lloyd08} showed that entangled states outperform unentangled states in a single-photon level to detect a low-reflectivity target under weak background noise. The performance was improved with coherent states \cite{SL09} and even more with two-mode squeezed vacuum (TMSV) states \cite{Tan08}, by means of QCB.
Since the proposal of the QI with TMSV state, there were several studies on QI with two-mode entangled states \cite{Sanz, Karsa, Lee, Prabhu21, Noh, Lee24, Pannu}, where the performance evaluation is related to the QCB. 
The scenario was extended to QI with multi-mode entangled states whose performance was evaluated with QCB \cite{Jung}, signal-to-noise ratio \cite{Gallego}, the decay rate of error probability under asymmetric quantum channel discrimination\cite{DePalma18}.

In contrast to the previous approaches, we are interested in the lower bound of symmetric quantum channel discrimination, i.e., the HB for fundamental understanding.
The HB can demonstrate the performance of quantum channel discrimination \cite{Pirandola} or quantum state discrimination \cite{Calsa10,Bae15}.
It can be rigorously studied under single-shot detection, leading to exact analytic solutions of fundamental detection limit.
A single-shot detection limit was theoretically studied in QI with two-qudit states, where a maximally entangled bipartite state is the optimal state under white noise environment \cite{Yung20}. For two-qubit cases, it was experimentally implemented\cite{Xu21}.
Here, we extend the two-qudit scenario to multi-qudit scenario, along with a question: Is a maximally entangled multipartite state the optimal state in the multi-qudit scenario? 
We study multi-qudit states in the QI under white noise environment. We obtain that the optimal state is not a maximally entangled multipartite state but a combination of maximally entangled bipartite state, while the multimode states can outperform the two-mode states in the QI.

\begin{figure}[h]
    \centering
    \includegraphics[width=10cm]{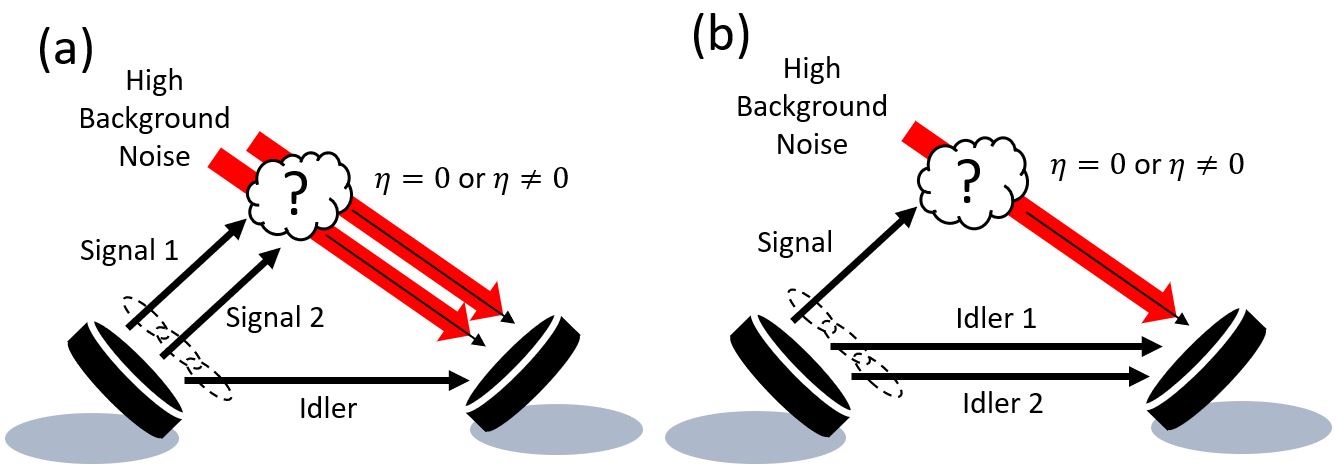}
    \includegraphics[width=5cm]{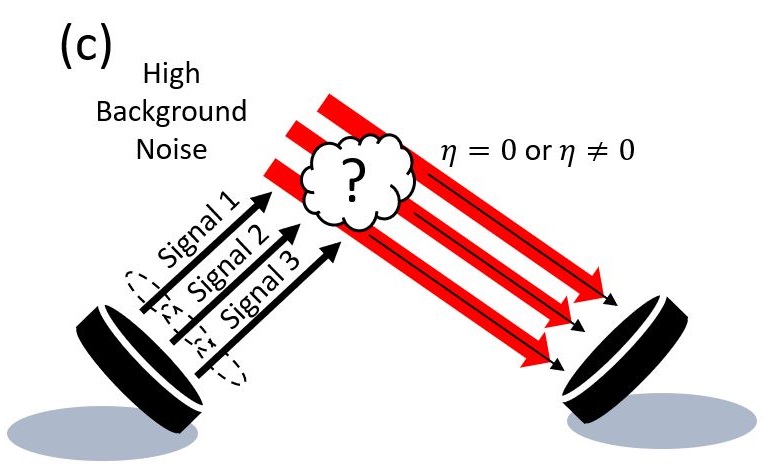}
    \caption{A schematic of QI with three modes in order to determine whether there is a target $(\eta\neq0)$ or not $(\eta=0)$, where $\eta$ is a target reflectivity. There are three configurations: (a) two-signal and one-idler ($2S1I$) modes , (b) one-signal and two-idler ($1S2I$) modes, and (c) three-signal ($3S$) modes. The modes in the dashed circles can be entangled or not.
  A separable state is the classical benchmark in the scenario.}
\label{Fig1}
\end{figure}

In Fig.~\ref{Fig1}, three-mode states can be classified with three configurations, such as (i) two-signal and one-idler, (ii) one-signal and two-idler, and (iii) three-signal. For each configuration, there are two types of genuine entangled state, namely Greenberger--Horne--Zeilinger (GHZ) state \cite{GHSZ} and W-state \cite{DVC00}, as well as bipartite entangled or totally separable states. 
By analyzing the fundamental limits of the states for each configuration, we can have positive results from entanglement between signal and idler qub(d)its whereas negative results from entanglement between different signal qub(d)its. 
It elucidates what kind of entanglement structure contributes to quantum advantage in QI. Additionally, we evaluate the quantum mutual information for each state and compare it with the detection limits, resulting in the same order of the performance in QI with three-qubit states. 
We extend the configuration to four-qudit states to figure out if the performance maintains the same order in both the HB and quantum mutual information.

\section*{Results}

\subsection*{Preliminaries}

We consider the following four topics as preliminaries: Three-qubit states, Helstrom bound, Holevo information, and fundamental detection limit of quantum illumination with two-qudit.

\subsubsection*{Three-qubit states}

For each configuration, we start with three-qubit states which are well classified into GHZ state, W-state, bipartite entangled states, and product states. The classification is based on the criterion that two quantum states belong to the same entanglement class if they can be converted into one another by stochastic local operations and classical communication (SLOCC) \cite{DVC00}. 
The last two classes are not considered as genuine tripartite entanglements, since they contain at least one separable qubit.
The GHZ state and W-state cannot be converted into one another by SLOCC, as they represent distinct classes of genuine tripartite entangled states in which no qubit has zero local entropy. It is also known that these are the only two types of genuine tripartite entanglement for three-qubit states \cite{DVC00}.

\begin{figure}[ht]
    \centering
    \includegraphics[width=12cm]{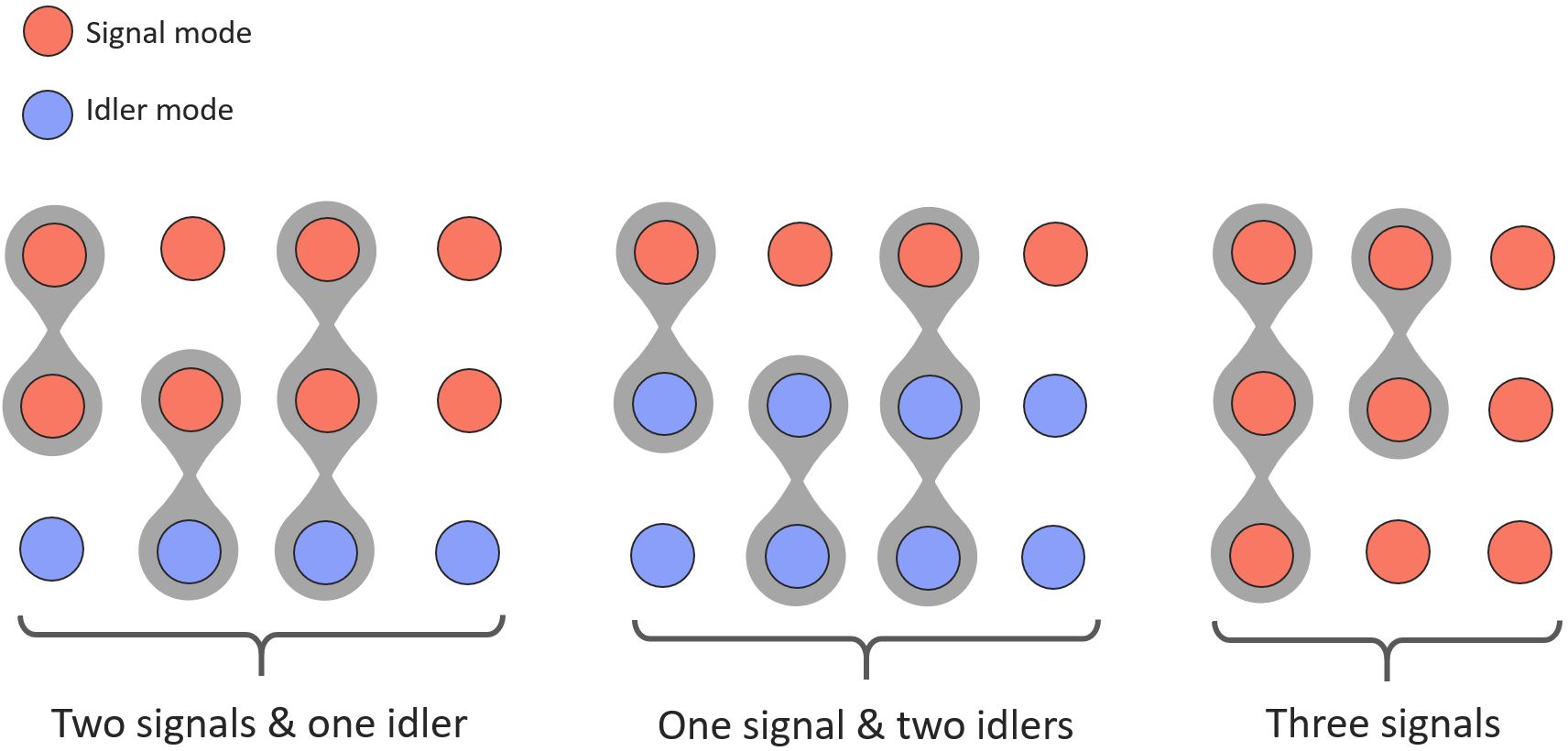}
    \caption{Three configurations with three-qubit states.  
    Red (blue) circle represents signal (idler) qubit. 
    For each configuration, entangled qubits are demonstrated with gray color.
    }
\label{Fig2}
\end{figure}

All the possible configurations are given in Fig.~\ref{Fig2}. 
For instance, employing the four classes of states, we consider a configuration with two-signal and one-idler qubits. 
The possible preparations are described as: (i) product state, (ii) signal-signal entangled state, (iii) signal-idler entangled state, (iv) GHZ state, and (v) W-state. To simplify the notation for describing the states, we use the hyphen symbol, -,  to indicate the absence of entanglement between two qubits.
The S-S-I state represents a product state, where no qubits are entangled. 
The SS-I state represents a signal-signal entangled state, where the two-signal qubits are only entangled with each other. 
The S-SI state represents a signal-idler entangled state, where only one signal is entangled with the idler qubit. 
The SSI state represents GHZ and W-states, where all the qubits are entangled with each other.
This notation is consistently used throughout this paper.

\subsubsection*{Helstrom bound} 
QI is to discriminate the presence and absence of a target with entangled states. Based upon a two-outcome POVM (Positive Operator-Valued Measure), we can ambiguously discriminate the two cases with minimum error. One measurement detects the presence of the target ($\rho_1$) while the other measurement detects the absence of the target ($\rho_0$). If a measurement does not fit into the presence (or absence) of the target, it produces errors, such as a false-alarm error and a miss-detection error. The sum of the errors is minimized over all possible POVMs, leading to the HB \cite{Helstrom} that is the lower bound of the detection error probability.

\begin{figure}[h!]
    \centering
    \includegraphics[width=12cm]{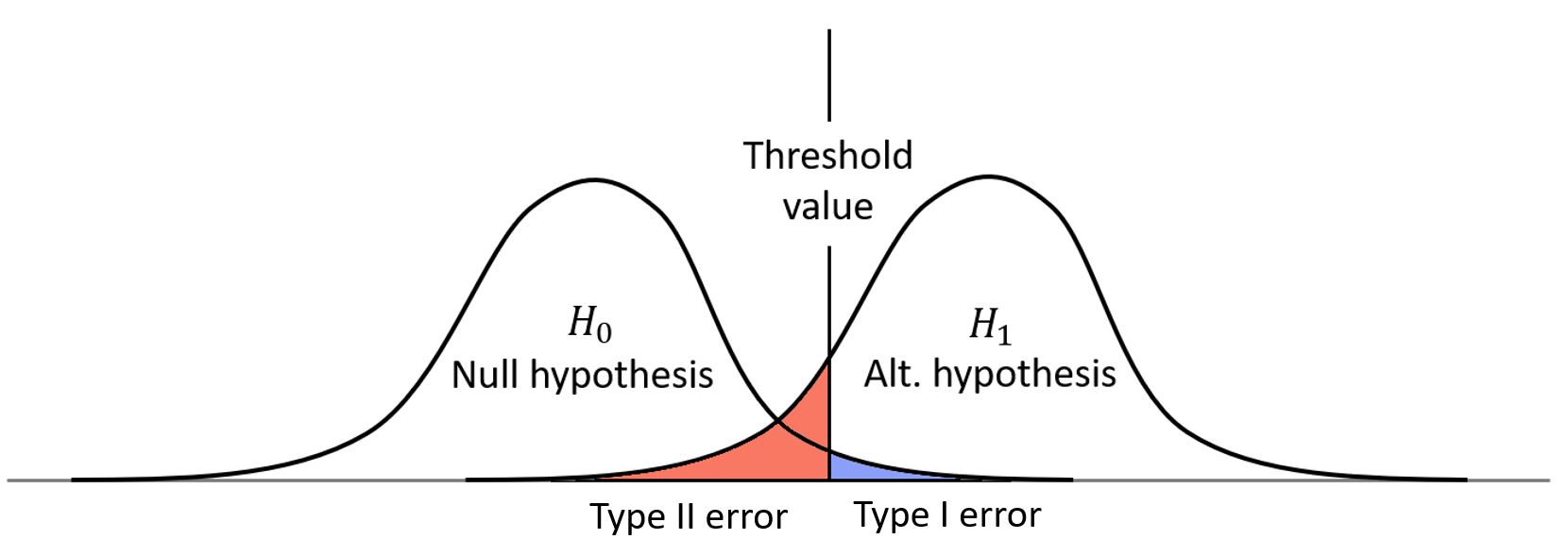}
    \caption{Hypothesis testing involving null hypothesis $(H_0)$ and alternative hypothesis $(H_1)$. 
    The probability of detecting $H_1$ when $H_0$ is actually true is called false-alarm probability (Type I error), and the probability of detecting $H_0$ when $H_1$ is actually true is called miss-detection probability (Type II error).
    The horizontal axis represents a measurable parameter, and the vertical axis represents a probability.}
    \label{Fig3}
\end{figure}

We explain the derivation of a single-shot detection limit with the HB.
Let the measurement of a receiver be $\left\{\Pi_0, \Pi_1 \right\}$. The states $\rho_0$ and $\rho_1$ are prepared with corresponding prior probabilities $p_0$ and $p_1$, respectively. Then, we can write the detection error probability of this measurement as
\begin{equation}
P_{err} = p_0 \mathrm{Tr}\left[ \Pi_1\rho_0\right] + p_1 \mathrm{Tr}\left[ \Pi_0\rho_1 \right],
\end{equation}
where $p_0 \mathrm{Tr}\left[ \Pi_1 \rho_0\right]$ corresponds to the false-alarm probability, and $p_1 \mathrm{Tr}\left[ \Pi_0\rho_1 \right]$ corresponds to the miss-detection probability, as shown in in Fig.~\ref{Fig3}. Using the property $\Pi_0 + \Pi_1 = I$, $P_{err}$ can be formulated as
\begin{eqnarray}
    P_{err} = \frac{1}{2} - \mathrm{Tr}\left[(p_1 \rho_1 - p_0 \rho_0)(\Pi_1 - \Pi_0)\right].
\end{eqnarray}
The optimal POVM is given by the $\Pi_1$ which is a projector onto the positive support of $(p_1 \rho_1 - p_0 \rho_0)$, and by the $\Pi_0$ which is a projector onto the negative support. Thus, the minimum detection error probability is given as:
\begin{equation}
P_{err} = \frac{1}{2}\left(1 - \lVert p_1 \rho_1 - p_0 \rho_0 \rVert\right),
\end{equation}
where $\lVert \sigma \rVert = \mathrm{Tr}(\sqrt{\sigma^{\dagger}\sigma})$ denotes the trace norm. This theorem is a powerful tool for investigating the fundamental limit of hypothesis testing to discriminate between two quantum states. Not only does it provide the lower bound of the detection error probability, but it also specifies the measurement required to achieve that limit.

\subsubsection*{Holevo information}
We introduce how to evaluate quantum mutual information as a measure of informational advantage in QI.
However, it is a challenge to maximize the mutual information over all possible POVMs, such that we can take an upper bound of the mutual information, leading to Holevo information.
Within a framework of quantum communication \cite{Weedbrook}, the Holevo information is analyzed as below.

Let $X = \left\{x,p_x\right\}$ be a random variable where $x$ is binary. Alice possesses this random variable, and when $x = 0$ with probability $p_0$, she sends the state $\rho_0$ to Bob. Otherwise, she sends $\rho_1$. Bob then employs a POVM $\mathcal{M} = \left\{\Pi_0,\Pi_1\right\}$ to discriminate between the received states $\rho_0$ and $\rho_1$. Let $K^{\mathcal{M}}$ denote Bob's measurement outcome, which is also a random variable. Then, we can define the mutual information $I(X,K^{\mathcal{M}})$ which quantifies the information shared between Alice and Bob by Bob's measurement $\mathcal{M}$.
The accessible information is defined as the maximum mutual information over all possible POVMs, i.e., $I_{acc} = \underset{\mathcal{M}}{\max}~I(X,K^{\mathcal{M}})$. This quantity can be used as a measure of the quantum advantage of QI compared to conventional illumination (CI). 
The Holevo information is represented by
\begin{eqnarray}
    \chi = S\left(\sum\limits_i p_i\rho_i\right)-\sum\limits_i p_i S(\rho_i), 
\end{eqnarray}
where $S(\rho)$ is the von Neumann entropy. If the states $\rho_0$ and $\rho_1$ commute, i.e.,$\left[\rho_0, \rho_1\right] = 0$, it is known that the accessible information $I_{acc}$ saturates the Holevo information $\chi$. 
We use the Holevo information to determine whether informational advantage guarantees performances of QI with multi-qudit states.

\subsubsection*{Fundamental detection limit of quantum illumination with two-qudit}
As a benchmark work, we briefly review the single-shot detection limit for QI using two-qudit states.
Previously, M.-H. Yung \textit{et al.} \cite{Yung20} studied the scenario under random noise of $\rho_E = \sum\limits_{i=1}^d \lambda_i \ket{\theta_i}_S\bra{\theta_i}$, where $\theta_i$ is a frequency mode. They showed that the HB can be expressed as a function of the prior probability $p_0$ and the target's reflectivity $\eta$.

For QI with two-qudit states, the output states are given by $\rho_0 = \rho_E \otimes \mathrm{Tr}_{S} (\ket{\psi}_{SI}\bra{\psi})$ and
    $\rho_1 = \eta \ket{\psi}_{SI}\bra{\psi} + (1-\eta) \rho_E \otimes \mathrm{Tr}_{S}(\ket{\psi}_{SI}\bra{\psi})$.
The corresponding HB is derived as: 
\begin{equation}
    P_{err} = 
    \begin{cases} 
        p_0 & \text{if } \gamma \geq 0 \\
        p_0+\gamma(1-\lambda_h) & \text{if } \gamma < 0 , \frac{p_1\eta}{|\gamma|} \geq \lambda_h\\
        p_1 & \text{if } \gamma < 0 , \frac{p_1\eta}{|\gamma|} < \lambda_h 
    \end{cases}
\end{equation}
where $\gamma = p_1(1-\eta)-p_0$ and $\lambda_h = \left(\sum\limits_i \frac{1}{\lambda_i}\right)^{-1}$. 
In certain regions where the detection error probability becomes $P_{err} = p_0$ or $p_1$, it indicates that a naive guess without any measurement achieves the minimum detection error probability. They termed it the non-illuminable region since no measurement provides better information than a naive guess based on the prior probability. For example, when $\gamma \geq 0$, the optimal POVM becomes $\Pi_1 =I, \Pi_0 = 0$, meaning that directly guessing the presence of the target yields the optimal error probability $p_0$. 
In contrast to the non-illuminable regions, the remaining parameter space is termed the illuminable region. In the illuminable region, the error probability is expressed in closed form using $p_0$ and $\eta$, achieving a value less than $\min(p_0,p_1)$, thus making measurement useful. The optimal POVM to achieve this limit is given by $\Pi_1 = \ket{\psi}_{SI}\bra{\psi}, \Pi_0 = I-\ket{\psi}_{SI}\bra{\psi}$. 

For CI with single-mode qudits, the output states are given by 
    $\rho_0 = \rho_E$ and $\rho_1 = \eta \ket{\psi}_S\bra{\psi} + (1-\eta) \rho_E$. 
The corresponding HB is derived as: 
\begin{equation}
    P_{err} = 
    \begin{cases} 
        p_0 & \text{if } \gamma \geq 0 \\
        p_0+\gamma(1-\lambda_{min}) & \text{if } \gamma < 0 , \frac{p_1\eta}{|\gamma|} \geq \lambda_{min}\\
        p_1 & \text{if } \gamma < 0 , \frac{p_1\eta}{|\gamma|} < \lambda_{min} 
    \end{cases}
\end{equation}
where the optimal POVM to achieve the limit on the illuminable region is given by $\Pi_1 = \ket{\psi}_{S}\bra{\psi}, \Pi_0 = I-\ket{\psi}_{S}\bra{\psi}$, representing the same form as the POVM used in QI.

Since $\lambda_{min} \geq \lambda_h = \left(\sum\limits_i \frac{1}{\lambda_i}\right)^{-1}$, 
QI takes a larger illuminable region than CI while the corresponding HB is lower in QI than in CI.
Under white noise, it also shows that the optimal state is a maximally entangled state and the quantum advantage is explained by quantum correlation measure.

\subsection*{QI with multi-qudit states}
In a single-shot detection limit of QI under white noise, we can extend an input bipartite state to a multipartite state and raise questions:
Does a maximally entangled multipartite state present the best peformance?
Is the quantum advantage explained by quantum correlation measure?
To answer the questions, we investigate three configurations with HB and Holevo information. 
Note that a white noise is given by $\rho_E = \frac{1}{d}\sum\limits_{n=0}^{d-1} \ket{n}_S\bra{n}$, where $d$ is the number of dimension.
Before taking them rigorously, we describe each configuration briefly as follows, which brings us some intuition.

For the two-signal and one-idler ($2S1I$) configuration, there are five possible states: (i) product state, (ii) signal-signal entangled state, (iii) signal-idler entangled state, (iv) GHZ state, and (v) W-state. When signals are lost, we receives only background noises in the signal mode. When signals interact with a target, signals are independently reflected from a target with a probability of $\eta$.  The corresponding states of $\rho_0$ and $\rho_1$ are formulated as:

\begin{eqnarray}
\rho_0 &=& \rho_{E_1}\otimes\rho_{E_2}\otimes \mathrm{Tr}_{S_{1} S_{2}}(\ket{\psi}_{S_1S_2I}\bra{\psi}),\nonumber\\ 
\rho_1 &=& \eta^2\ket{\psi}_{S_1S_2I}\bra{\psi}+(1-\eta)\eta\{\rho_{E_1}\otimes \mathrm{Tr}_{S_1}(\ket{\psi}_{S_1S_2I}\bra{\psi}) +\rho_{E_2}\otimes \mathrm{Tr}_{S_2}(\ket{\psi}_{S_1S_2I}\bra{\psi}) \} \nonumber\\
&&+(1-\eta)^2\rho_{E_1}\otimes\rho_{E_2}\otimes \mathrm{Tr}_{S_{1} S_{2}}(\ket{\psi}_{S_1S_2I}\bra{\psi}).
\label{eq:2S1I conf}
\end{eqnarray}
The terms of $\eta(1-\eta)$ break entanglement between two signals, resulting in a mixture. Thus, a maximally entangled multipartite state cannot present the best performance in the $2S1I$ configuration.

For the one-signal and two-idler ($1S2I$) configuration, in the above five possible states, we have an idler-idler entangled state instead of the signal-signal entangled state. After interacting with a target or not, the corresponding states are formulated as:
\begin{eqnarray}
&\rho_0 = \rho_{E_1} \otimes \mathrm{Tr}_S(\ket{\psi}_{SI_1I_2}\bra{\psi}), 
&\rho_1 = \eta\ket{\psi}_{SI_1I_2}\bra{\psi}+(1-\eta)\rho_{E_1}\otimes \mathrm{Tr}_S(\ket{\psi}_{SI_1I_2}\bra{\psi}).
\label{eq:1S2I conf}
\end{eqnarray}
Since idler states do not interact with a target, an idler-idler entanglement does not contribute to the performance. 
From now on, thus, we do not consider the idler-idler entanglement.

For the three-signal ($3S$) configuration, there are four possible states: product state $(\text{S-S-S})$, bipartite entangled state $(\text{SS-S})$, GHZ state, and W-state. After interacting with a target, the corresponding states are formulated as:
\begin{eqnarray}
&\rho_0 = \rho_{E_1}\otimes\rho_{E_2}\otimes\rho_{E_3},
&\rho_1 = \sum\limits_{i=0}^{3} \eta^{3-i} (1-\eta)^i\bigg[\underset{U_j \in \binom{\left\{1,2,3\right\}}{i}}{\sum} \bigotimes_{j_k \in U_j} \rho_{E_{j_{k}}} \otimes \mathrm{Tr}_{S_{j_1}...S_{j_k}} (\ket{\psi}_{S_1S_2S_3}\bra{\psi})\bigg].
\label{eq:3S conf}
\end{eqnarray}
Entanglement among signals is broken down by interacting with the target, resulting in a mixture, so that the corresponding performance can be worse than the product state $(\text{S-S-S})$.

\subsubsection*{Helstrom bound}

For different multimode states, we cannot directly compare error probabilities since there are multiple illuminable regions with different boundaries. For example, it is not guaranteed that a specific point of $(\eta, p_0)$ belongs to the same region for different multimode states. Each region belongs to a different optimal measurement, such that it can lead to an unfair evaluation to directly compare the error probabilities across different states. Thus, we consider mean value of HBs.

We investigate HB of QI with three-qubit states under white noise environment. Let $\Pi_i$ represent the measurement corresponding to $\rho_i$. When $\Pi_0$ is the projector onto the negative support of $(p_1\rho_1-p_0\rho_0)$ and $\Pi_1$ is the projector onto the positive support of $(p_1\rho_1-p_0\rho_0)$, they become the optimal POVM. In the context of QI, the eigenvalues of $(p_1\rho_1-p_0\rho_0)$ depend on the prior probability $p_0$ and the reflectivity $\eta$. Consequently, the optimal POVM $\left\{\Pi_0,\Pi_1\right\}$ changes with $p_0$ and $\eta$. Based on this criterion, we can divide the parameter space $\left(p_0, \eta \right) \in \left[0,1\right]\times\left[0,1\right]$ into distinct regions which have different POVMs, respectively. 

\begin{figure}[ht!]
    \centering
    \includegraphics[width=8cm]{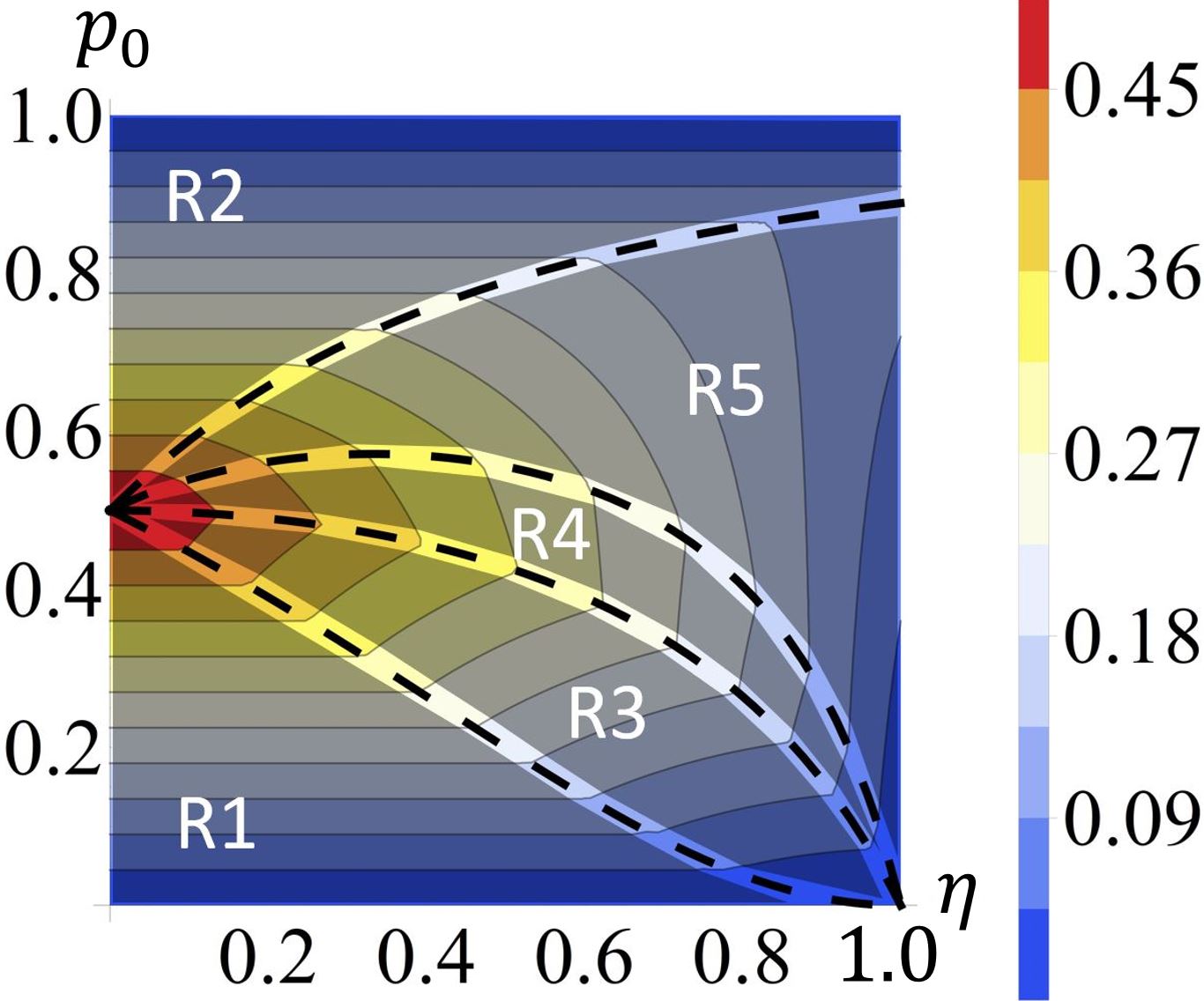}
    \caption{HB of \text{S-SI} state in $2S1I$ configuration as a function of $\eta$ and $p_0$. 
    The dashed lines represent the boundaries of the HB.
    R1 and R2 represent non-illuminable regions, whereas R3, R4, and R5 represent illuminable regions.}
    \label{Fig5}
\end{figure}

In Fig.~\ref{Fig5}, for example, the HB for the \(\text{S-SI}\) state in the $2S1I$ configuration has five regions in the parameter space. Region 1 requires the POVM \(\left\{\Pi_0, \Pi_1\right\} = \left\{0, I\right\}\), and region 2 requires the POVM \(\left\{\Pi_0, \Pi_1\right\} = \left\{I, 0\right\}\). These are non-illuminable regions, analogous to the non-illuminable regions observed in QI with two-qudit \cite{Yung20}. Regions from $3$ to $5$ require nontrivial and distinct POVMs, resulting in strictly lower error probabilities than \(\min(p_0, p_1)\). These regions are termed illuminable regions. 
QI with three-qubit generally exhibits multiple illuminable regions.
In the two-qudit case, where there is only one illuminable region, it is sufficient to analyze the advantage by comparing the analytical error probabilities within that region. 
However, in the three-qubit case, it is infeasible to directly compare error probabilities due to the existence of multiple illuminable regions with different boundaries.
Thus, alternatively, we consider the mean HB over the parameter space $\left( p_0, \eta \right) \in \left[0,1\right]\times\left[0,1\right]$ as a measure of performance by integrating the HB over this region. \\

\begin{figure}[ht!]
    \centering
    \includegraphics[width=12cm]{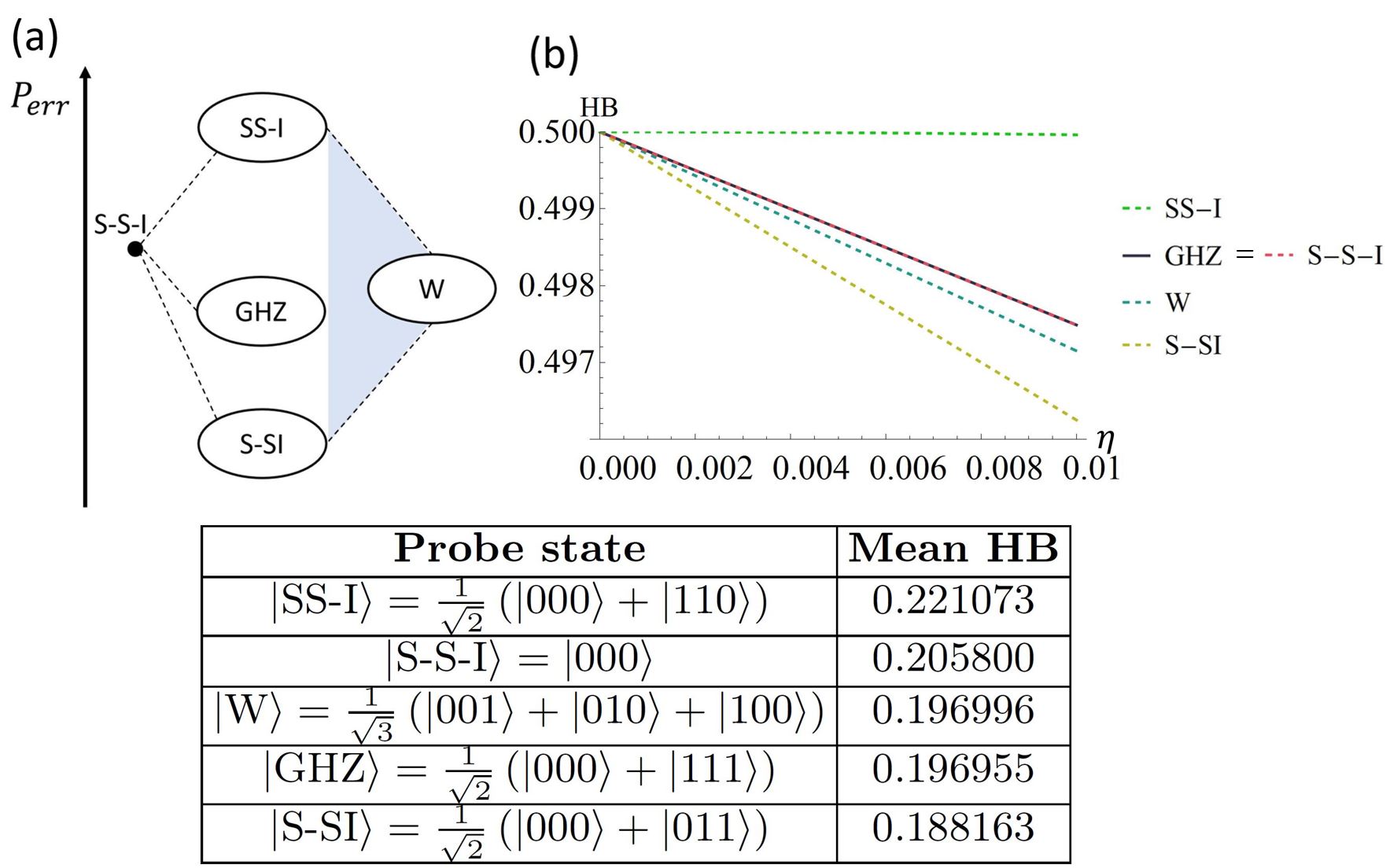}
    \caption{
    (a) Mean HB for $2S1I$ configuration, which is summarized in a table. SS-I, GHZ, and S-SI states can be converted to S-S-I state by a single parameter (dashed line). 
    W-state has detection error probability between the S-SI state and the SS-I state with two parameters (blue zone). 
    (b) HB as a function of $\eta \in [0,0.01]$ at $p_0=0.5$, where the order of the HB is S-SI<W<GHZ=S-S-I<SS-I.}
    \label{Fig6}
\end{figure}

\textbf{Configuration 1: Two signals and One idler}\\
From the states $\rho_0$ and $\rho_1$ of Eq.~(\ref{eq:2S1I conf}), we consider five possible states $\ket{\psi}_{SSI}$ to find the minimum HB. 
First, there are four possible states as follows:
\begin{eqnarray}
    &&\ket{\text{GHZ}}_{SSI} = \cos(\theta/2)\ket{000}_{SSI} + \sin(\theta/2)\ket{111}_{SSI},
    ~\ket{\text{S-SI}}_{SSI} = \cos(\theta/2)\ket{000}_{SSI} + \sin(\theta/2)\ket{011}_{SSI},\nonumber\\
    &&\ket{\text{SS-I}}_{SSI} = \cos(\theta/2)\ket{000}_{SSI} + \sin(\theta/2)\ket{110}_{SSI},
    ~\ket{\text{S-S-I}}_{SSI} = \ket{000}_{SSI}.
    \label{eq:2S1I states}
\end{eqnarray} 
Since the eigenvalues of $(p_1\rho_1-p_0\rho_0)$ is independent of the relative phase, the states are only parametrized with $\theta \in [0,\pi]$.
The corresponding mean HB is also a continuous function of superposition ratio $\theta$. 
Second, we consider another type of class, namely the W class. A more generalized version of the W-state is written as:
\begin{eqnarray}
    \ket{\text{W}}_{SSI} = x_1\ket{001}_{SSI}+x_2\ket{010}_{SSI}+x_3\ket{100}_{SSI},
    \label{eq:2S1I W-state}
\end{eqnarray}
where $x_1^2+x_2^2+x_3^2 = 1$. Unlike the states of Eq.~(\ref{eq:2S1I states}) that are parameterized by a single parameter, the state of Eq.~(\ref{eq:2S1I W-state}) is parameterized with two parameters. 
We obtain explicit HBs as a function of $\eta$ and $p_0$ for some representative states for each class. We choose optimal values of parameters, such as $\theta = \pi/2$ in the states of Eq.~(\ref{eq:2S1I states}) and $x_1=x_2=x_3=1/\sqrt{3}$ in the state of Eq.~(\ref{eq:2S1I W-state}).
Through numerical analysis, we find that the state which minimizes the mean HB is $\ket{\text{S-SI}}_{SSI} = \frac{1}{\sqrt{2}}\ket{0}_{S_i}(\ket{01}_{S_j I}+\ket{10}_{S_j I})$ $(i,j=1,2,~i\neq j)$, where the signal and idler qubits are maximally entangled whereas another signal qubit is separable from the other qubits. Conversely, the state which maximizes the mean HB is $\ket{\text{SS-I}}_{SSI} = \frac{1}{\sqrt{2}}(\ket{01}_{S_1 S_2}+\ket{10}_{S_1 S_2})\ket{0}_{I}$, where the signal qubits are maximally entangled whereas the idler qubit is separable from the other qubits.
In Fig.~\ref{Fig6}, we present an ordering of the mean HBs among the different classes with a simple diagram and a table.
 Moreover, we look into an interesting range of $p_0=0.5$ and $\eta\leq 0.01$ which represent an unknown prior probability and a low-reflectivity target.  Compared to the mean HB of Fig.~\ref{Fig6} (a), we obtain the same optimal state of the HB in Fig.~\ref{Fig6} (b), whereas the GHZ state presents the same performance as the S-S-I state which presents lower performance than the W state. 

In Methods, we provide the detailed derivation of analytic solution of HB which contains the information about the boundary, minimum detection error probability, and optimal POVM.
In $2S1I$ configuration, the order of the mean HB among different states demonstrates that entanglement between signal qubits degrades the performance, whereas entanglement between signal and idler qubits enhances it. Based on the linear entropy of quantum state bipartitions, we provide an intuitive explanation in Methods. \\


\begin{figure}[ht]
    \centering
    \includegraphics[width=12cm]{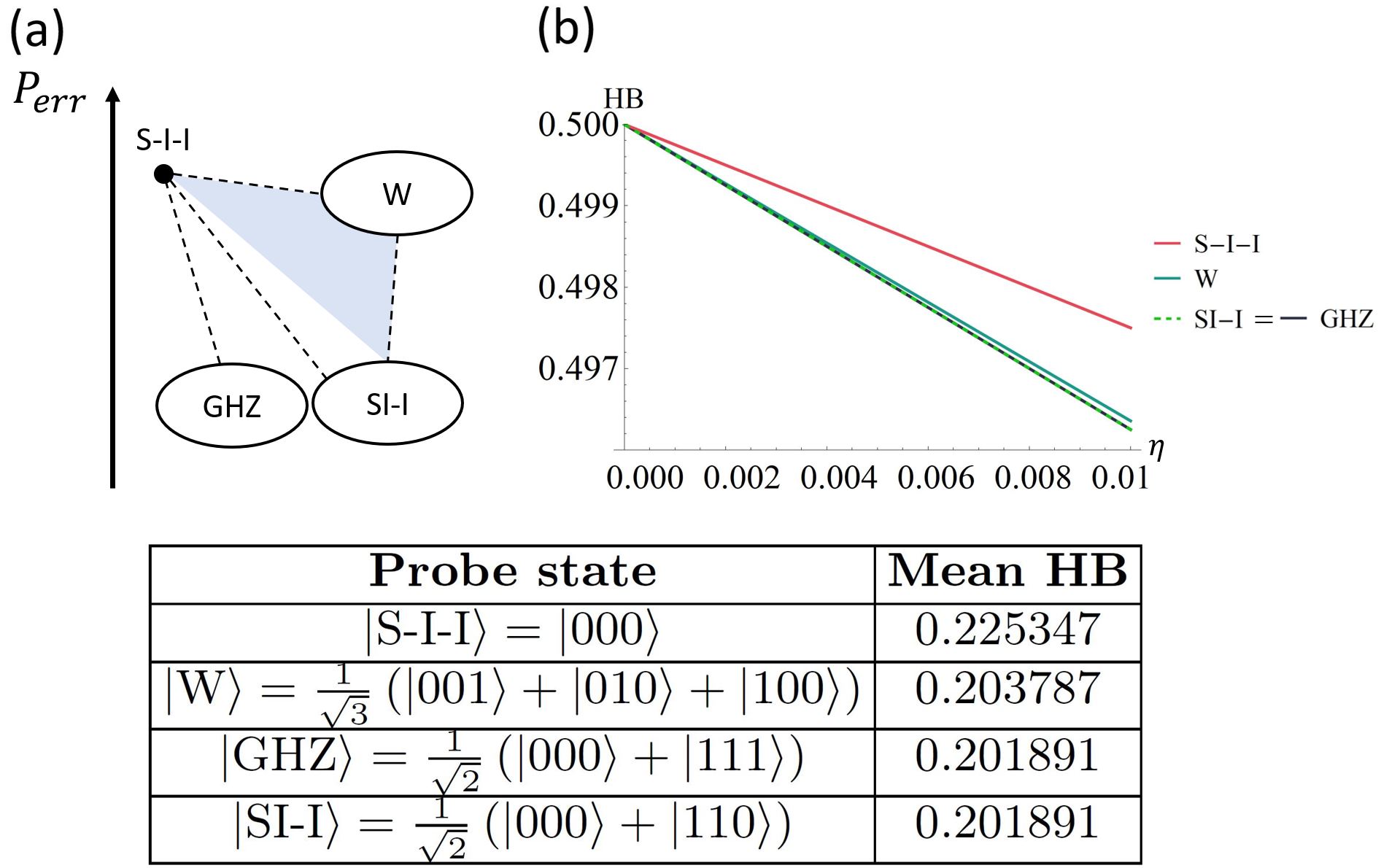}
    \caption{(a) Mean HB for $1S2I$ configuration, which is summarized in a table. (b) HB as a function of $\eta \in [0,0.01]$ at $p_0=0.5$, where the order of the HB is SI-I=GHZ<W<S-S-I.}
    \label{1S2I class}
\end{figure}

\textbf{Configuration 2: One signal and Two idlers}\\
From the states $\rho_0$ and $\rho_1$ of Eq.~(\ref{eq:1S2I conf}), we consider the following states:
\begin{align}
    \begin{split}
    \ket{\text{GHZ}}_{SII} &= \cos(\theta/2)\ket{000}_{SII} + \sin(\theta/2)\ket{111}_{SII}, 
    ~\ket{\text{S-II}}_{SII} = \cos(\theta/2)\ket{000}_{SII} + \sin(\theta/2)\ket{011}_{SII},\\
    \ket{\text{SI-I}}_{SII} &= \cos(\theta/2)\ket{000}_{SII} + \sin(\theta/2)\ket{110}_{SII},
    ~\ket{\text{W}}_{SII} = x_1\ket{001}_{SII}+x_2\ket{010}_{SII}+x_3\ket{100}_{SII},\\
    \ket{\text{S-I-I}}_{SII} &= \ket{000}_{SII}, 
    \end{split}
\end{align}
where the relative phase is also ignored as $2S1I$ configuration. 
Each state can be continuously converted to $\ket{\text{S-I-I}}_{SII} = \ket{000}_{SII}$ as $\theta \rightarrow 0$ except W class. 
We choose optimal values of the parameters, such as $\theta = \pi/2$ in the S-II, GHZ, S-I-I, SI-I classes and $x_1=x_2=x_3=1/\sqrt{3}$ in the W class.
Through numerical analysis, we find that the state which minimizes the mean HB is $\ket{\text{SI-I}}_{SII} = \frac{1}{\sqrt{2}}(\ket{01}_{SI}+\ket{10}_{SI})\ket{0}_{I}$, where the signal and idler qubits are maximally entangled whereas another idler qubit is separable from the other qubits. Since the separable idler qubit does not contribute to the performance, it presents the same performance as a maximally bipartite state as well as the GHZ state. Conversely, the state which maximizes the mean HB is $\ket{\text{S-I-I}}_{SII}= \ket{000}_{SII}$, where all the qubits are separable.
In Fig.~\ref{1S2I class}, we present the ordering of the mean HB among the different states with a simple diagram and a table.
Moreover, we look into an interesting range of $p_0=0.5$ and $\eta\leq 0.01$.  Compared to the mean HB of Fig.~\ref{1S2I class} (a), we obtain the same ordering of the HB in Fig.~\ref{1S2I class} (b). \\

\begin{figure}[ht]
    \centering
    \includegraphics[width=12cm]{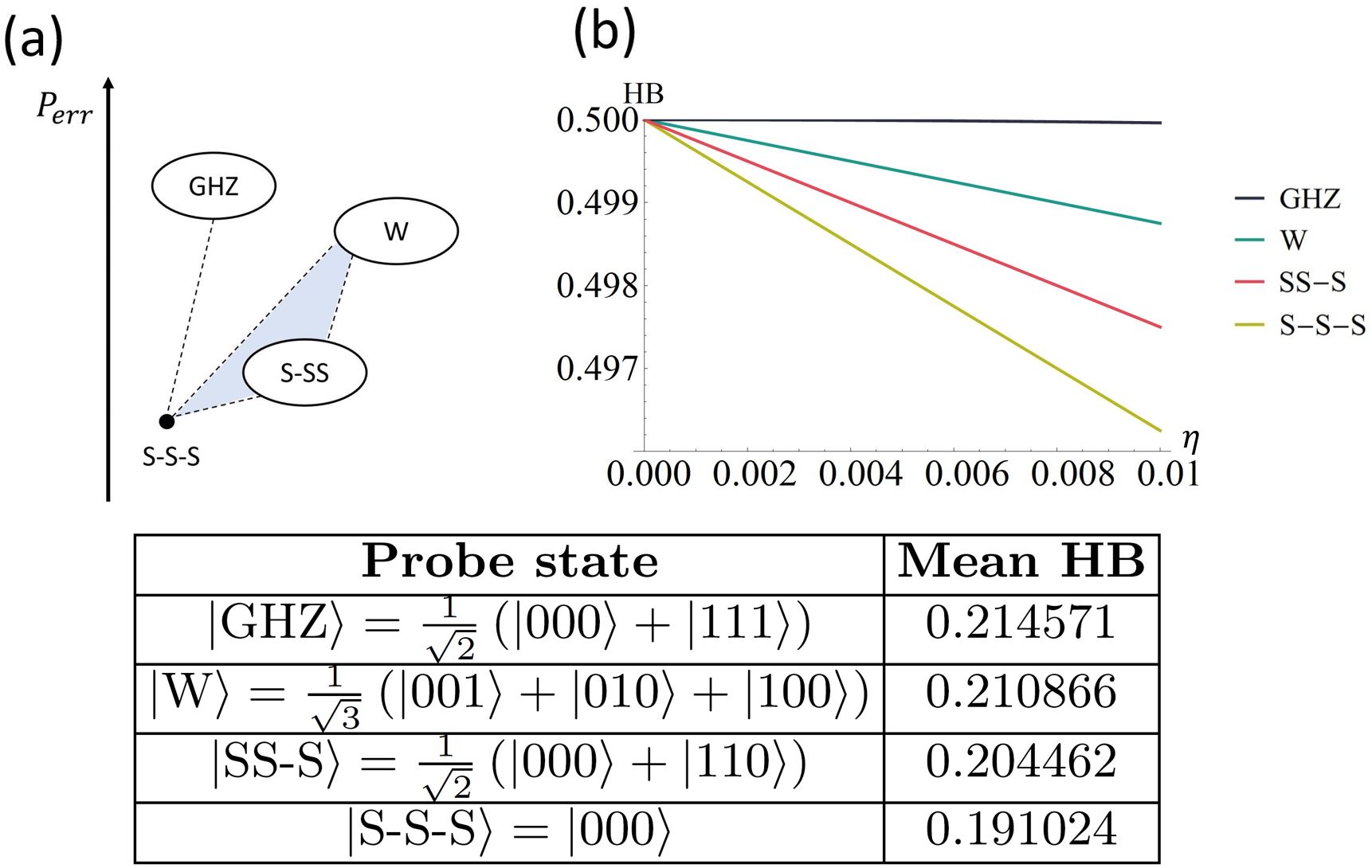}
    \caption{(a) Mean HB for $3S$ configuration, which is summarized in a table.  (b) HB as a function of $\eta \in [0,0.01]$ at $p_0=0.5$, where the order of the HB is S-S-S<SS-S<W<GHZ.}
    \label{3S class}
\end{figure}

\textbf{Configuration 3: Three signals}\\
From the states $\rho_0$ and $\rho_1$ of Eq.~(\ref{eq:3S conf}), we consider the following states :

\begin{eqnarray}
 &&\ket{\text{W}}_{SSS} = x_1\ket{001}_{SSS}+x_2\ket{010}_{SSS}+x_3\ket{100}_{SSS},
 ~\ket{\text{GHZ}}_{SSS} = \cos(\theta/2)\ket{000}_{SSS} + \sin(\theta/2)\ket{111}_{SSS},\nonumber\\ 
 &&\ket{\text{SS-S}}_{SSS} = \cos(\theta/2)\ket{000}_{SSS} + \sin(\theta/2)\ket{110}_{SSS},
 ~\ket{\text{S-S-S}}_{SSS} = \ket{000}_{SSS},
    \label{eq:3S states}
\end{eqnarray}
where each state can be continuously converted to $\ket{\text{S-S-S}}_{SSS} = \ket{000}_{SSS}$ as $\theta \rightarrow 0$, except W class. 
We choose optimal values of the parameters, such as $\theta = \pi/2$ in the GHZ, SS-S, S-S-S classes and  $x_1=x_2=x_3=1/\sqrt{3}$ in the W class.
Through numerical analysis, we find that the state which minimizes the mean HB is a separable state $\ket{\text{S-S-S}}_{SSS}$. Conversely, the state which maximizes the mean HB is $\ket{\text{GHZ}}_{SSS}$.
In Fig.~\ref{3S class}, we present the ordering of the mean HB across the different states with a simple diagram and a table.
Moreover, we look into an interesting range of $p_0=0.5$ and $\eta\leq 0.01$.  Compared to the mean HB of Fig.~\ref{3S class} (a), we obtain the same ordering of the HB in Fig.~\ref{3S class} (b). 

 \begin{table}[ht]
     \centering
     \small
     \begin{tabular}{|>{\bfseries}c|c|c|c|c|}
         \hline
         \textbf{} & \textbf{1S1I} & \textbf{2S1I} & \textbf{1S2I} & \textbf{3S} \\ \hline
         \textbf{Separable}   & 0.225347     & 0.205800     & 0.225347   & 0.191024       \\ \hline
         \textbf{S-I entangled}   & 0.201891   &\cellcolor{red!30}0.188163    & 0.201891  & -        \\ \hline
         \textbf{S-S entangled}   & -   & 0.221073      & - & 0.204462     \\ \hline
         \textbf{I-I entangled}   & -    & -       & 0.225347    & -      \\ \hline
         \textbf{GHZ}   & -      & 0.196955    & 0.201891      & 0.214571       \\ \hline
         \textbf{W}   & -   & 0.196996       & 0.203787       & 0.210866      \\ \hline
     \end{tabular}
     \caption{Mean HBs for QI with three-qubit, including two-qubit case ($1S1I$). The best probe state is the \text{S-SI} state in a red box.}
     \label{tab:Mean HB for three qubit QI}
 \end{table}

In table~\ref{tab:Mean HB for three qubit QI}, we summarize the mean HBs, compared to the case of QI with two-qubit. 
For QI with three qubits, the optimal probe state is not a maximally tripartite entangled state but a pair of a signal-idler entangled state with a signal state, i.e., $\ket{\text{S-SI}} = \frac{1}{\sqrt{2}}\ket{0}_{S_i}\left(\ket{00}_{S_jI}+\ket{11}_{S_jI}\right)(i,j = 1,2, i \neq j)$, where it outperforms the two-qubit case because one more signal qubit is sent to a target. Moreover, we obtain that the performance gets worse by a signal-signal entanglement since the signal-signal entanglement is broken to become a mixture, resulting in being worse performance than a product state. In a specific range of $p_0=0.5$ and $\eta\leq 0.01$, the ordering of the mean HBs is partly confirmed, with no difference between \text{S-SI} state and \text{SI} state.


\subsubsection*{Holevo information}
For two-qubit cases, the quantum advantage of QI is explained by the quantum mutual information of the output state \cite{Weedbrook}. 
For multi-qudit states, similarly, we compare the order of mean HB with the order of mean quantum mutual information to determine whether the informational advantage consistently guarantees a low-detection error probability. Using Holevo information that is an upper bound of quantum mutual information, we obtain that high quantum mutual information of an output state guarantees low-detection error probability for the three-qubit, three-qutrit, and four-qubit states, but this relation does not consistently hold for four-ququart states.\\

\indent\textbf{Quantum illumination with three-qubit}\\
Holevo information is computable when density operators commute as $[\rho_0,\rho_1] = 0$.
Among five possible states that we consider, however, W-state does not satisfy the commutation relation so that it cannot be computed with Holevo information. 

Similar to the HB, Holevo information is also a function of $p_0$ and $\eta$. To compare different states, we evaluate the average value of Holevo information by integrating it over the parameter space. For QI with three qubits, we find that mean Holevo information and the mean HB are well aligned in Table~\ref{tab:three qubit Holevo}. In other words, the state with high mean Holevo information corresponds to low-detection error probability in three-qubit scenarios. \\

\begin{table}[ht!]
    \centering
    \resizebox{\textwidth}{!}
    {\begin{tabular}{|c|c|c|c|c|c|c|c|c|c|c|}
        \hline
        \textbf{} & \textbf{S-SI} & \textbf{S-S-S} & \textbf{GHZ(2S1I)} & \textbf{SI-I} & \textbf{GHZ(1S2I)} & \textbf{SS-S} & \textbf{S-S-I} & \textbf{GHZ(3S)} & \textbf{SS-I} & \textbf{S-I-I}\\
        \hline
        \textbf{Helstrom} & 0.188163 & 0.191024 & 0.196955 & 0.201891 & 0.201891 & 0.204462  & 0.205800 & 0.214571 & 0.221073 & 0.225347\\
        \hline
        \textbf{Holevo} & 0.0968226 & 0.0935365 & 0.0823021 & 0.0712934 & 0.0712934 & 0.0709855  & 0.0674483 & 0.0532995 & 0.0424548 & 0.0365761\\
        \hline
    \end{tabular}}
    \caption{Mean HB and mean Holevo information for QI with three-qubit. Different states are arranged in an increasing order of mean HB as well as a decreasing order of mean Holevo information, from left to right.}
    \label{tab:three qubit Holevo}
\end{table}

\indent\textbf{Quantum illumination for multi-dimension and multi-partition}\\
We extend our analysis to multiple dimensions and multiple partitions in order to examine whether the ordering of mean HB is consistently guaranteed by mean Holevo information. 
We evaluate all possible configurations for QI with three qutrits, four qubits, and four ququarts. Then, we identify a case of four ququarts in which the ordering of mean Holevo information violates the ordering of the mean HB. The complete set of comparisons is provided in Methods. 

A full classification of entanglement for such states remains an open problem; therefore, we focus on several well-known types of states, including generalized GHZ states, cyclic states, and partially entangled states.
A cyclic state is one of high-dimensional multipartite entangled states. It was investigated theoretically \cite{Cabello,Jo, Jo25} and experimentally \cite{Pan} in high-dimensional quantum teleportation. In a \(d\)-partite and \(d\)-dimensional system, a cyclic state is defined as:
\begin{equation}
    \ket{d-\Psi} = \frac{1}{\sqrt{d!}}\text{perm}(\Lambda_d)\ket{vac},
\end{equation}
where \(\text{perm}(A)\) is the permanent of the matrix \(A\), and \(\ket{vac}\) is a vacuum state. The matrix \(\Lambda_d\) is defined as:
\begin{equation}
\Lambda_d = 
    \begin{pmatrix}
      \hat{a}^{\dagger}_{00} & \hat{a}^{\dagger}_{01} & \dots & \hat{a}^{\dagger}_{0(d-1)}\\
      \hat{a}^{\dagger}_{10} & \hat{a}^{\dagger}_{11} & \dots & \hat{a}^{\dagger}_{1(d-1)}\\
      \vdots & \vdots & \ddots & \vdots \\
      \hat{a}^{\dagger}_{(d-1)0} & \hat{a}^{\dagger}_{(d-1)1} & \dots & \hat{a}^{\dagger}_{(d-1)(d-1)}\\
    \end{pmatrix},
\end{equation}
where \(\hat{a}^{\dagger}_{ij}\) denotes the creation operator of which orthonormal mode and path label are $i$ and $j$, respectively \cite{Jo}.  
This type of state satisfies \([\rho_0, \rho_1] = 0\), allowing us to measure quantum mutual information by computing the Holevo information.  

\begin{table}[ht!]
    \centering
    \begin{tabular}{|c|c|c|c|c|c|c|c|c|}
        \hline
        \textbf{} & \textbf{S-S-SI} & \textbf{S-SSI} & \textbf{SS-SI} & \textbf{GHZ} & \textbf{S-S-S-I} & \textbf{Cyclic} & \textbf{S-SS-I} & \textbf{SSS-I}\\
        \hline
        \textbf{Helstrom} & 0.14128 & 0.15310 & \cellcolor{red!30}0.15559 & \cellcolor{red!30}0.15666 & 0.15711 & 0.16500 & 0.17753 & 0.19828 \\
        \hline
        \textbf{Holevo} & 0.19014 & 0.16892 & \cellcolor{red!30}0.16081 & \cellcolor{red!30}0.16215 & 0.15852 & 0.14686 & 0.12198 & 0.08794 \\
        \hline
    \end{tabular}
    \caption{Mean HB and mean Holevo information for $3S1I$. Different states are arranged in an increasing order of mean HB as well as a decreasing order of mean Holevo information, from left to right. The red boxes indicate that the mean Holevo information is inconsistent with the order of the mean HB.}
    \label{tab:3S1I mean HB and mean Holveo bound}
\end{table}



In Table~\ref{tab:3S1I mean HB and mean Holveo bound}, we show the performance of QI with four-ququarts, in a configuration of three-signal and one-idler modes.
We find a case of that high mean Holevo information does not guarantee low-detection error probability in the red-colored boxes. Specifically, the \(\text{SS-SI}\) state has less mean Holevo information compared to the \(\text{GHZ}\) state. 
However, the \(\text{SS-SI}\) state provides a lower detection error probability than the \(\text{GHZ}\) state, as indicated by the mean HB. Thus, for the multi-qudit cases, Holevo information does not always capture the relationship between quantum correlation and the detection error probability. In Methods, we show mean Holevo information and mean HB for other scenarios .

\section*{Discussion}
We studied single-shot detection limits for QI with three-qubit (or qutrit) and four-qubit (or ququart) states, which was analyzed with the mean HB under white noise environment. 
We obtained that the performance is enhanced by entanglement between signal and idler qub(d)its whereas the performance is degraded by entanglement between signal qub(d)its. The signals independently interact with a target, such that the signal-signal entanglement is broken to become a mixture, leading to worse performance than a product state.
Through a comprehensive comparison across all states and configurations, we identified that the optimal state is not a maximally entangled multipartite state but a combination of maximally entangled bipartite state. For three-qub(d)it states, the best performance is obtained by a signal-idler entangled state with a signal state ($\text{S-SI}$). For four-qub(d)it states, it is obtained by two pairs of signal-idler entangled states ($\text{SI-SI}$).

The HBs that we obtained are explained by quantum correlation of the output states, which is derived by Holevo information, i.e., an upper bound of quantum mutual information.
We showed that the Holevo information guarantees the order of detection error probabilities in QI with three-qubit, three-qutrit, and four-qubit states. However, the consistency broke in the case of four-ququart states. It implies that quantum correlation measure cannot always be a measure of advantage in QI under white noise environment.

An intriguing question for future research would be to determine the optimal state under random noise environment, where the noise state is not identically distributed. For a general $N$-qudit system, additionally, it would be an important topic of investigating whether we can still hold the disadvantage from signal-signal entanglement and the advantage from signal-idler entanglement. These directions would help to refine our understanding of the role of entanglement in QI and expand the scope of its applications. Furthermore, it could be studied with variational quantum algorithms for QI \cite{Sai}.

\section*{Methods}

\subsection*{Deriving analytic solutions of Helstrom bound}
For two-signal and one-idler configuration, we provide detailed derivation of explicit HB for representative states of different types of entanglement.
HB is given as $P_{err} = \frac{1}{2}\left(1-\lVert p_1\rho_1-p_0\rho_0\rVert\right)$.

If $\gamma_1 = p_1(1-\eta)^2-p_0\geq0$, ($p_1\rho_1-p_0\rho_0$) is nonnegative, so
    $P_{err} = \frac{1}{2}(1-(p_1-p_0)) = p_0$,
and POVM is given as $\Pi_1 = I, \Pi_0 = 0$. This region is called region 1, which is a non-illuminable region.
If $\gamma_1<0$, at least one eigenvalue of ($p_1\rho_1-p_0\rho_0$) is negative. Then $P_{err}$ is summarized as : 

\begin{align}
    \begin{split}
    (p_1\rho_1-p_0\rho_0) =&~ \gamma_1\{\rho_{E_1}\otimes\rho_{E_2}\otimes\textrm{Tr}_{S_1S_2}(\ket{\psi}_{S_1S_2I}\bra{\psi})
    -\alpha_1\{\rho_{E_1}\otimes\mathrm{Tr}_{S_1}(\ket{\psi}_{S_1S_2I}\bra{\psi}) +\rho_{E_2}\otimes \mathrm{Tr}_{S_2}(\ket{\psi}_{S_1S_2I}\bra{\psi})\}\\
    &-\alpha_2\ket{\psi}_{S_1S_2I}\bra{\psi}\},
    \end{split}
\end{align}
where $\alpha_1 = \frac{p_1\eta(1-\eta)}{-\gamma_1}$, $\alpha_2 = \frac{p_1\eta^2}{-\gamma_1}$. Since $\gamma_1<0$, $\alpha_1>0$ and $\alpha_2>0$. Depending on the states, we obtain the different eigenvalues of $\Omega_{\text{2S1I}} = \rho_{E_1}\otimes\rho_{E_2}\otimes\textrm{Tr}_{S_1S_2}(\ket{\psi}_{S_1S_2I}\bra{\psi})-\alpha_1\{\rho_{E_1}\otimes\mathrm{Tr}_{S_1}(\ket{\psi}_{S_1S_2I}\bra{\psi}) +\rho_{E_2}\otimes \mathrm{Tr}_{S_2}(\ket{\psi}_{S_1S_2I}\bra{\psi})\}-\alpha_2\ket{\psi}_{S_1S_2I}\bra{\psi}$.

\subsubsection*{S-S-I state}
The representative state for the S-S-I class is $\ket{\text{S-S-I}} = \ket{000}$. The eigenvalues of $\Omega_{\text{2S1I}}$ are :
\begin{equation}
    0,0,0,0,\frac{1}{4},\frac{1}{4}-\frac{\alpha_1}{2},\frac{1}{4}-\frac{\alpha_1}{2},\frac{1}{4}-\alpha_1-\alpha_2.
\end{equation}
The region is divided as follows.

\textbf{Region 3} : $\gamma_1<0$, $0>\frac{1}{4}-\frac{\alpha_1}{2}$. The minimal detection error probability is given by : 
\begin{equation}
    P_{err} = \frac{1}{2}\left(1-(-\gamma_1)(2\alpha_1+\alpha_2-\frac{1}{2})\right).
\end{equation} 
The optimal POVM is given by 
    $\Pi_1 = \ket{000}\bra{000}+\ket{010}\bra{010}+\ket{100}\bra{100}$.

\textbf{Region 4} : $\gamma_1<0$, $\frac{1}{4}-\frac{\alpha_1}{2}\geq0>\frac{1}{4}-\alpha_1-\alpha_2$. The minimal detection error probability is given by : 
\begin{equation}
    P_{err} = \frac{1}{2}\left(1-(-\gamma_1)(\alpha_2+\frac{1}{2})\right).
\end{equation} 
The optimal POVM is given by 
    $\Pi_1 = \ket{000}\bra{000}$.

\textbf{Region 2} :$\gamma_1<0$, $\frac{1}{4}-\alpha_1-\alpha_2\geq0$. The minimal detection error probability is given by : 
\begin{equation}
    P_{err} = p_1.
\end{equation} 
The optimal POVM is given by  
    $\Pi_1 = 0$.

\subsubsection*{GHZ state}
The representative state for the GHZ class is $\ket{\text{GHZ}} = \frac{1}{\sqrt{2}}\left(\ket{000}+\ket{111}\right)$. The eigenvalues of $\Omega_{\text{2S1I}}$ are :
\begin{equation}
    \frac{1}{8},\frac{1}{8},\frac{1}{8}-\frac{\alpha_1}{2},\frac{1}{8}-\frac{\alpha_1}{4},\frac{1}{8}-\frac{\alpha_1}{4},\frac{1}{8}-\frac{\alpha_1}{4},\frac{1}{8}-\frac{\alpha_1}{4},\frac{1}{8}-\frac{\alpha_1}{2}-\alpha_2.
\end{equation}
The region is divided as follows.

\textbf{Region 3} : $\gamma_1<0$, $0>\frac{1}{8}-\frac{\alpha_1}{4}$. The minimal detection error probability is given by : 
\begin{equation}
    P_{err} = \frac{1}{2}\left(1-(-\gamma_1)(2\alpha_1+\alpha_2-\frac{1}{2})\right).
\end{equation} 
The optimal POVM is given by 
    $\Pi_1 = \ket{\text{GHZ}}\bra{\text{GHZ}}+\ket{\phi_1}\bra{\phi_1}+\ket{010}\bra{010}+\ket{011}\bra{011}+\ket{100}\bra{100}+\ket{101}\bra{101}$,
where $\ket{\phi_1} = \frac{1}{\sqrt{2}}\left(\ket{000}+\ket{111}\right)$.

\textbf{Region 4} : $\gamma_1<0$, $\frac{1}{8}-\frac{\alpha_1}{4}\geq0>\frac{1}{8}-\frac{\alpha_1}{2}$. The minimal detection error probability is given by : 
\begin{equation}
    P_{err} = \frac{1}{2}\left(1-(-\gamma_1)(\alpha_2+\frac{1}{2})\right).
\end{equation} 
The optimal POVM is given by
    $\Pi_1 = \ket{\text{GHZ}}\bra{\text{GHZ}}+\ket{\phi_1}\bra{\phi_1}$.

\textbf{Region 5} : $\gamma_1<0$, $\frac{1}{8}-\frac{\alpha_1}{2}\geq0>\frac{1}{8}-\frac{\alpha_1}{2}-\alpha_2$. The minimal detection error probability is given by : 
\begin{equation}
    P_{err} = \frac{1}{2}\left(1-(-\gamma_1)(-\alpha_1+\alpha_2+\frac{3}{4})\right).
\end{equation} 
The optimal POVM is given by 
    $\Pi_1 = \ket{\text{GHZ}}\bra{\text{GHZ}}$.

\textbf{Region 2} :$\gamma_1<0$, $\frac{1}{8}-\frac{\alpha_1}{2}-\alpha_2\geq0$. The minimal detection error probability is given by : 
\begin{equation}
    P_{err} = p_1.
\end{equation} 
The optimal POVM is given by 
    $\Pi_1 = 0$.

\subsubsection*{W-state}
The representative state for the W class is $\ket{\text{W}} = \frac{1}{\sqrt{3}}\left(\ket{001}+\ket{010}+\ket{100}\right)$. The eigenvalues of $\Omega_{\text{2S1I}}$ are :
\begin{align}
    \begin{split}
    \frac{1}{12},\frac{1}{12}-\frac{\alpha_1}{6},\frac{1}{6}-\frac{\alpha_1}{3},\frac{1}{6}-\frac{\alpha_1}{3},
    \frac{3-6\alpha_1\pm\sqrt{1-4\alpha_1+36\alpha_1^2}}{24},
    \frac{3-8\alpha_1-12\alpha_2\pm\sqrt{1+32\alpha_1^2-8\alpha_2+128\alpha_1\alpha_2+144\alpha_2^2}}{24}.
    \end{split}
\end{align}

The region is divided as follows.

\textbf{Region 3} : $\gamma_1<0$, $0>\frac{3-8\alpha_1-12\alpha_2-\sqrt{1+32\alpha_1^2-8\alpha_2+128\alpha_1\alpha_2+144\alpha_2^2}}{24}$. The minimal detection error probability is given by : 
\begin{equation}
    P_{err} = \frac{1}{2}\left(1-(-\gamma_1)\left(\frac{-7+18\alpha_1+12\alpha_2+\sqrt{1-4\alpha_1+36\alpha_1^2}}{12}\right)\right).
\end{equation} 

The optimal POVM is given by : 
\begin{align}
    \begin{split}
    \Pi_1 = \ket{\phi_1}\bra{\phi_1}+\ket{\phi_2}\bra{\phi_2}+\ket{\phi_3}\bra{\phi_3}+\ket{\phi_4(\eta,p_0)}\bra{\phi_4(\eta,p_0)}+
    \ket{\phi_5(\eta,p_0)}\bra{\phi_5(\eta,p_0)}+\ket{\phi_6(\eta,p_0)}\bra{\phi_6(\eta,p_0)},
    \end{split}
\end{align}
where 
\begin{align}
    \begin{split}
    &\ket{\phi_1} = \frac{1}{\sqrt{2}}\left(\ket{011}-\ket{101}\right), \ket{\phi_2} = \frac{1}{\sqrt{2}}\left(\ket{010}-\ket{100}\right), \ket{\phi_3} = \ket{000}, \\
    &\ket{\phi_4(\eta,p_0)} = \frac{4\alpha_1}{\sqrt{32\alpha_1^2+\left(-1+2\alpha_1+\sqrt{1-4\alpha_1+36\alpha_1^2}\right)^2}}(\ket{011}+\ket{101})+\\
    &\frac{-1+2\alpha_1+\sqrt{1-4\alpha_1+36\alpha_1^2}}{\sqrt{32\alpha_1^2+\left(-1+2\alpha_1+\sqrt{1-4\alpha_1+36\alpha_1^2}\right)^2}}\ket{110}, \\
    &\ket{\phi_5(\eta,p_0)} = \frac{1-4\alpha_2+\sqrt{1+32\alpha_1^2-8\alpha_2+128\alpha_1\alpha_2+144\alpha_2^2}}{\sqrt{32(\alpha_1+2\alpha_2)^2+\left(1-4\alpha_2+\sqrt{1+32\alpha_1^2-8\alpha_2+128\alpha_1\alpha_2+144\alpha_2^2}\right)^2}}\ket{001}+\\
    &\frac{4(\alpha_1+2\alpha_2)}{\sqrt{32(\alpha_1+2\alpha_2)^2+\left(1-4\alpha_2+\sqrt{1+32\alpha_1^2-8\alpha_2+128\alpha_1\alpha_2+144\alpha_2^2}\right)^2}}(\ket{010}+\ket{100}),\\
    &\ket{\phi_6(\eta,p_0)} = \frac{1-4\alpha_2-\sqrt{1+32\alpha_1^2-8\alpha_2+128\alpha_1\alpha_2+144\alpha_2^2}}{\sqrt{32(\alpha_1+2\alpha_2)^2+\left(-1+4\alpha_2+\sqrt{1+32\alpha_1^2-8\alpha_2+128\alpha_1\alpha_2+144\alpha_2^2}\right)^2}}\ket{001}+\\
    &\frac{4(\alpha_1+2\alpha_2)}{\sqrt{32(\alpha_1+2\alpha_2)^2+\left(-1+4\alpha_2+\sqrt{1+32\alpha_1^2-8\alpha_2+128\alpha_1\alpha_2+144\alpha_2^2}\right)^2}}(\ket{010}+\ket{100}).
    \end{split}
\end{align}
\textbf{Region 4} : $\gamma_1<0$, $\frac{3-8\alpha_1-12\alpha_2-\sqrt{1+32\alpha_1^2-8\alpha_2+128\alpha_1\alpha_2+144\alpha_2^2}}{24}\geq0>\frac{1}{6}-\frac{\alpha_1}{3}$. The minimal detection error probability is given by : 
\begin{equation}
    P_{err} = \frac{1}{2}\left(1-(-\gamma_1)\left(\frac{-4+10\alpha_1+\sqrt{1-4\alpha_1+36\alpha_1^2}+\sqrt{1+32\alpha_1^2-8\alpha_2+128\alpha_1\alpha_2+144\alpha_2^2}}{12}\right)\right).
\end{equation} 

The optimal POVM is given by : 
\begin{equation}
    \Pi_1 =  \ket{\phi_1}\bra{\phi_1}+\ket{\phi_2}\bra{\phi_2}+\ket{\phi_3}\bra{\phi_3}+\ket{\phi_4(\eta,p_0)}\bra{\phi_4(\eta,p_0)}+\ket{\phi_6(\eta,p_0)}\bra{\phi_6(\eta,p_0)}.
\end{equation}

\textbf{Region 5} : $\gamma_1<0$, $\frac{1}{6}-\frac{\alpha_1}{3}\geq0>\frac{3-6\alpha_1-\sqrt{1-4\alpha_1+36\alpha_1^2}}{24}$. The minimal  detection error probability is given by : 
\begin{equation}
    P_{err} = \frac{1}{2}\left(1-(-\gamma_1)\left(\frac{6-10\alpha_1+\sqrt{1-4\alpha_1+36\alpha_1^2}+\sqrt{1+32\alpha_1^2-8\alpha_2+128\alpha_1\alpha_2+144\alpha_2^2}}{12}\right)\right).
\end{equation} 

The optimal POVM is given by 
    $\Pi_1 = \ket{\phi_4(\eta,p_0)}\bra{\phi_4(\eta,p_0)}+\ket{\phi_6(\eta,p_0)}\bra{\phi_6(\eta,p_0)}$.

\textbf{Region 6} : $\gamma_1<0$, $\frac{3-6\alpha_1-\sqrt{1-4\alpha_1+36\alpha_1^2}}{24}\geq0>\frac{3-8\alpha_1-12\alpha_2-\sqrt{1+32\alpha_1^2-8\alpha_2+128\alpha_1\alpha_2+144\alpha_2^2}}{24}$. The minimal detection error probability is given by : 
\begin{equation}
    P_{err} = \frac{1}{2}\left(1-(-\gamma_1)\left(\frac{9-16\alpha_1+\sqrt{1+32\alpha_1^2-8\alpha_2+128\alpha_1\alpha_2+144\alpha_2^2}}{12}\right)\right).
\end{equation} 
The optimal POVM is given by 
    $\Pi_1 = \ket{\phi_6(\eta,p_0)}\bra{\phi_6(\eta,p_0)}$.

\textbf{Region 2} :$\gamma_1<0$, $\frac{3-8\alpha_1-12\alpha_2-\sqrt{1+32\alpha_1^2-8\alpha_2+128\alpha_1\alpha_2+144\alpha_2^2}}{24}\geq0$. The minimal detection error probability is given by : 
\begin{equation}
    P_{err} = p_1.
\end{equation} 
The optimal POVM is given by 
   $\Pi_1 = 0$.

For the W-state, the optimal POVM is not constant on each region but depends on the reflectivity of target $\eta$ and the prior probability $p_0$. As a result, achieving the fundamental detection limit requires precise information about $\eta$ and $p_0$, which is infeasible.

\subsubsection*{S-SI state}
The representative state for the S-SI class is $\ket{\text{S-SI}} = \frac{1}{\sqrt{2}}\left(\ket{000}+\ket{011}\right)$. The eigenvalues of $\Omega_{\text{2S1I}}$ are :
\begin{equation}
    \frac{1}{8},\frac{1}{8},\frac{1}{8},\frac{1}{8}-\frac{\alpha_1}{4},\frac{1}{8}-\frac{\alpha_1}{4},\frac{1}{8}-\frac{\alpha_1}{4},\frac{1}{8}-\frac{\alpha_1}{2},\frac{1}{8}-\frac{3\alpha_1}{4}-\alpha_2.
\end{equation}
The region is divided as follows.

\textbf{Region 3} : $\gamma_1<0$, $0>\frac{1}{8}-\frac{\alpha_1}{4}$. The minimal detection error probability is given by : 
\begin{equation}
    P_{err} = \frac{1}{2}\left(1-(-\gamma_1)(2\alpha_1+\alpha_2-\frac{1}{4})\right).
\end{equation} 
The optimal POVM is given by 
    $\Pi_1 = \ket{\phi_1}\bra{\phi_1}+\ket{\phi_2}\bra{\phi_2}+\ket{\text{S-SI}}\bra{\text{S-SI}}+\ket{001}\bra{001}+\ket{010}\bra{010}$,
where $\ket{\phi_1} = \frac{1}{\sqrt{2}}\left(\ket{100}+\ket{111}\right)$, $\ket{\phi_2} = \frac{1}{\sqrt{2}}\left(\ket{000}-\ket{011}\right)$.

\textbf{Region 4} : $\gamma_1<0$, $\frac{1}{8}-\frac{\alpha_1}{4}\geq0>\frac{1}{8}-\frac{\alpha_1}{2}$. The minimal detection error probability is given by : 
\begin{equation}
    P_{err} = \frac{1}{2}\left(1-(-\gamma_1)(\frac{\alpha_1}{2}+\alpha_2+\frac{1}{2})\right).
\end{equation} 
The optimal POVM is given by 
    $\Pi_1 = \ket{\phi_1}\bra{\phi_1}+\ket{\text{S-SI}}\bra{\text{S-SI}}$.

\textbf{Region 5} : $\gamma_1<0$, $\frac{1}{8}-\frac{\alpha_1}{2}\geq0>\frac{1}{8}-\frac{3\alpha_1}{4}-\alpha_2$. The minimal error probability is given by : 
\begin{equation}
    P_{err} = \frac{1}{2}\left(1-(-\gamma_1)(-\frac{\alpha_1}{2}+\alpha_2+\frac{3}{4})\right).
\end{equation} 
The optimal POVM is given by : 
\begin{equation}
    \Pi_1 = \ket{\text{S-SI}}\bra{\text{S-SI}}.
\end{equation}

\textbf{Region 2} :$\gamma_1<0$, $\frac{1}{8}-\frac{3\alpha_1}{4}-\alpha_2\geq0$. The minimal detection error probability is given by : 
\begin{equation}
    P_{err} = p_1.
\end{equation} 
The optimal POVM is given by 
    $\Pi_1 = 0$.

\begin{figure}[ht!]
    \centering
    \includegraphics[width=9cm]{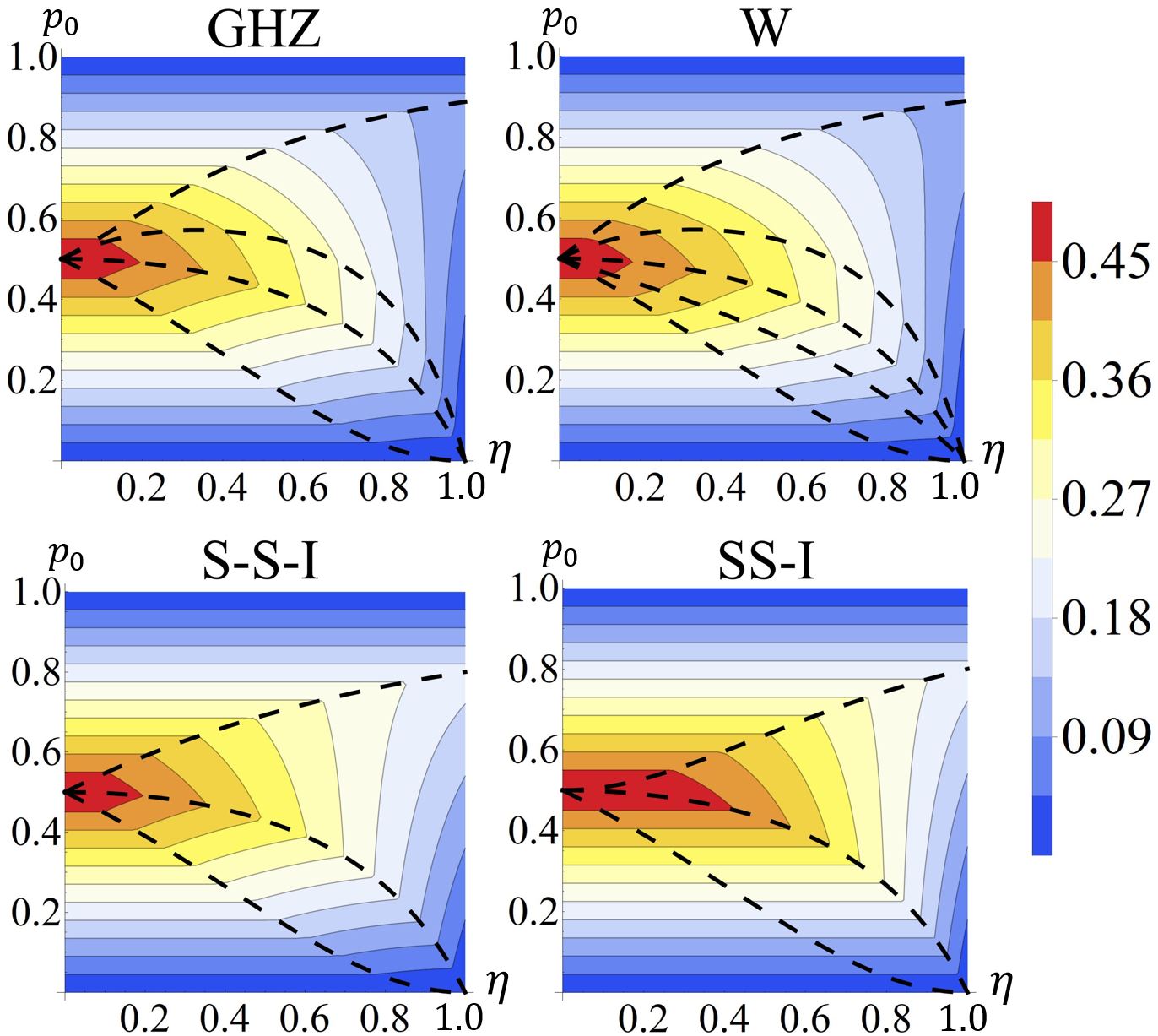}
    \caption{HBs of $2S1I$ as a function of $p_0$ and $\eta$. 
    The dashed lines represent the boundaries of the analytic solutions for the HBs.}
    \label{Fig7}
\end{figure}

\subsubsection*{SS-I state}
The representative state for the SS-I class is $\ket{\text{SS-I}} = \frac{1}{\sqrt{2}}\left(\ket{000}+\ket{110}\right)$. The eigenvalues of $\Omega_{\text{2S1I}}$ are :
\begin{equation}
    0,0,0,0,\frac{1}{4}-\frac{\alpha_1}{2},\frac{1}{4}-\frac{\alpha_1}{2},\frac{1}{4}-\frac{\alpha_1}{2},\frac{1}{4}-\frac{\alpha_1}{2}-\alpha_2.
\end{equation}
The region is divided as follows.

\textbf{Region 3} : $\gamma_1<0$, $0>\frac{1}{4}-\frac{\alpha_1}{2}$. The minimal detection error probability is given by : 
\begin{equation}
    P_{err} = \frac{1}{2}\left(1-(-\gamma_1)(-1+2\alpha_1+\alpha_2)\right).
\end{equation} 
The optimal POVM is given by 
    $\Pi_1 = \ket{\text{SS-I}}\bra{\text{SS-I}}+\ket{\phi_1}\bra{\phi_1}+\ket{010}\bra{010}+\ket{100}\bra{100}$,
where $\ket{\phi_1} = \frac{1}{\sqrt{2}}\left(\ket{000}-\ket{110}\right)$.

\textbf{Region 4} : $\gamma_1<0$, $\frac{1}{4}-\frac{\alpha_1}{2}\geq0>\frac{1}{4}-\frac{\alpha_1}{2}-\alpha_2$. The minimal detection error probability is given by : 
\begin{equation}
    P_{err} = \frac{1}{2}\left(1-(-\gamma_1)(-\alpha_1+\alpha_2+\frac{1}{2})\right).
\end{equation} 
The optimal POVM is given by
    $\Pi_1 = \ket{\text{SS-I}}\bra{\text{SS-I}}$.

\textbf{Region 2} :$\gamma_1<0$, $\frac{1}{4}-\frac{\alpha_1}{2}-\alpha_2\geq0$. The minimal detection error probability is given by : 
\begin{equation}
    P_{err} = p_1.
\end{equation} 
The optimal POVM is given by 
    $\Pi_1 = 0$.

The explicit HBs are drawn in Figs.~\ref{Fig5} and \ref{Fig7}.
In the same way, we can formulate the $\Omega_{\text{1S2I}}$ and $\Omega_{\text{3S}}$ and derive the analytic solution of HB by analyzing their eigenvalues for each state.  
Here, we present the analytic solutions of HBs for each configuration, which is evaluated with their respective optimal states.

\subsubsection*{1S2I configuration}
For the SI-I state, we provide an analytic solution of the boundary and a minimum detection error probability as follows : 

\textbf{Region 1} : $1-\frac{p_0}{p_1}\geq\eta$. The minimal error probability is given by 
    $P_{err} = p_0$ whose optimal POVM is given by $\Pi_1 = I$.

\textbf{Region 2} : $1-\frac{p_0}{p_1}<\eta$ and $p_0\geq1-\frac{1}{3\eta+2}$. The minimal error probability is given by 
    $P_{err} = p_1$ whose optimal POVM is given by $\Pi_1 = 0$.

\begin{figure}[ht!]
    \centering
    \includegraphics[width=9cm]{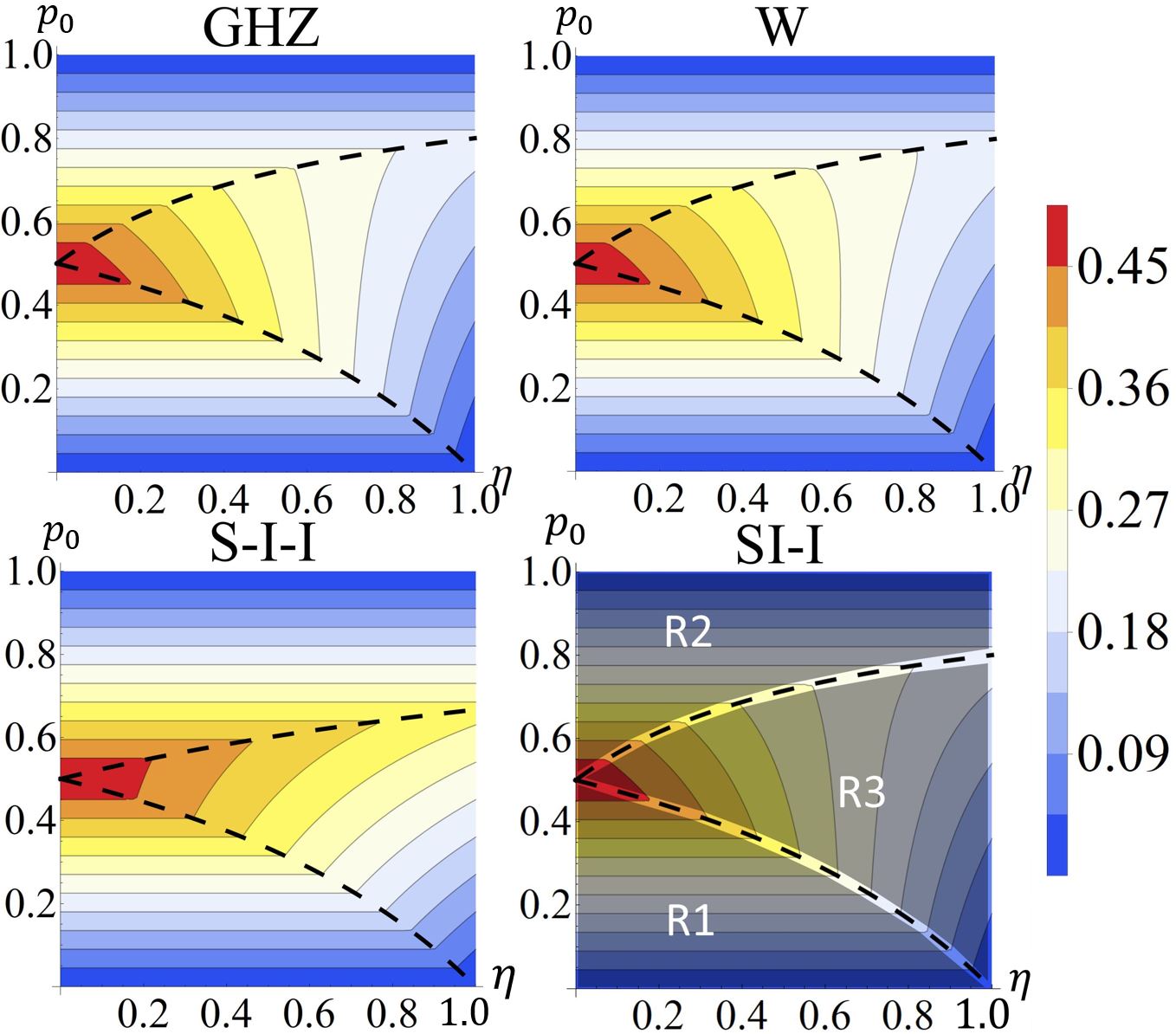}
    \caption{HBs of $1S2I$ as a function of $p_0$ and $\eta$.
    The dashed lines represent the boundaries of the analytic solutions for the HBs.}
    \label{1S2I HB}
\end{figure}

\textbf{Region 3} : $1-\frac{p_0}{p_1}<\eta$ and $1-\frac{1}{3\eta+2}>p_0$. The minimal error probability is given by 
\begin{equation}
    P_{err} = p_0+\frac{3}{4}\gamma_2,
\end{equation}
where $\gamma_2 = p_1(1-\eta)-p_0.$ We can observe that when $\gamma_2\geq0$, the parameters belong to region 1, whereas for $\gamma_2<0$, they belong to other regions. The optimal POVM is given by $\Pi_1= \ket{\psi}\bra{\psi}$, where $\ket{\psi}$ is $\ket{\text{SI-I}}$ or $\ket{\text{GHZ}}$. 
The HB of $\text{SI-I}$ and $\text{GHZ}$ states is the exactly same as the optimal HB of QI with two-qubit.
The explicit HBs are drawn in Fig. \ref{1S2I HB}.


\subsubsection*{3S configuration}
For the S-S-S state, we provide an analytic solution of the boundary and a minimum detection error probability as follows : 

\textbf{Region 1} : $1-\left(\frac{p_0}{p_1}\right)^{\frac{1}{3}}\geq\eta$. The minimal detection error probability is given by 
    $P_{err} = p_0$ whose optimal POVM is given by $\Pi_1 = I$.

\textbf{Region 2} : $1-\left(\frac{p_0}{p_1}\right)^{\frac{1}{3}}<\eta$ and $p_0\geq1-\frac{1}{\eta^3+3\eta^2+3\eta+2}$. The minimal detection error probability is given by 
    $P_{err} = p_1$ whose optimal POVM is given by $\Pi_1 = 0$.

\textbf{Region 3} : $1-\left(\frac{p_0}{p_1}\right)^{\frac{1}{3}}<\eta$ and $1-\frac{1}{\eta^3-\eta^2-\eta+2}>p_0$. The minimal detection error probability is given by 
\begin{equation}
    P_{err} = p_0+\frac{1}{8}\gamma_3,
\end{equation}
where $\gamma_3 = p_1(1-\eta)^3-p_0.$ We can observe that when $\gamma_3\geq0$, the parameters belong to region 1, whereas for $\gamma_3<0$, they belong to other regions. The optimal POVM is given by
$\Pi_1 = \ket{\psi}\bra{\psi}+I-\ket{000}\bra{000}-\ket{111}\bra{111}$,
where $\ket{\psi} = \ket{\text{S-S-S}}$.

\textbf{Region 4} : $1-\left(\frac{p_0}{p_1}\right)^{\frac{1}{3}}<\eta$ and $1-\frac{1}{-\eta^3-\eta^2+\eta+2}>p_0\geq1-\frac{1}{\eta^3-\eta^2-\eta+2}$. The minimal detection error probability is given by 
\begin{equation}
    P_{err} = \frac{1}{4}\left(2-3p_1\eta+p_1\eta^3\right),
\end{equation}
where the optimal POVM is given by
$\Pi_1 = \ket{\psi}\bra{\psi}+\ket{001}\bra{001}+\ket{010}\bra{010}+\ket{100}\bra{100}$.

\begin{figure}[ht!]
    \centering
    \includegraphics[width=9cm]{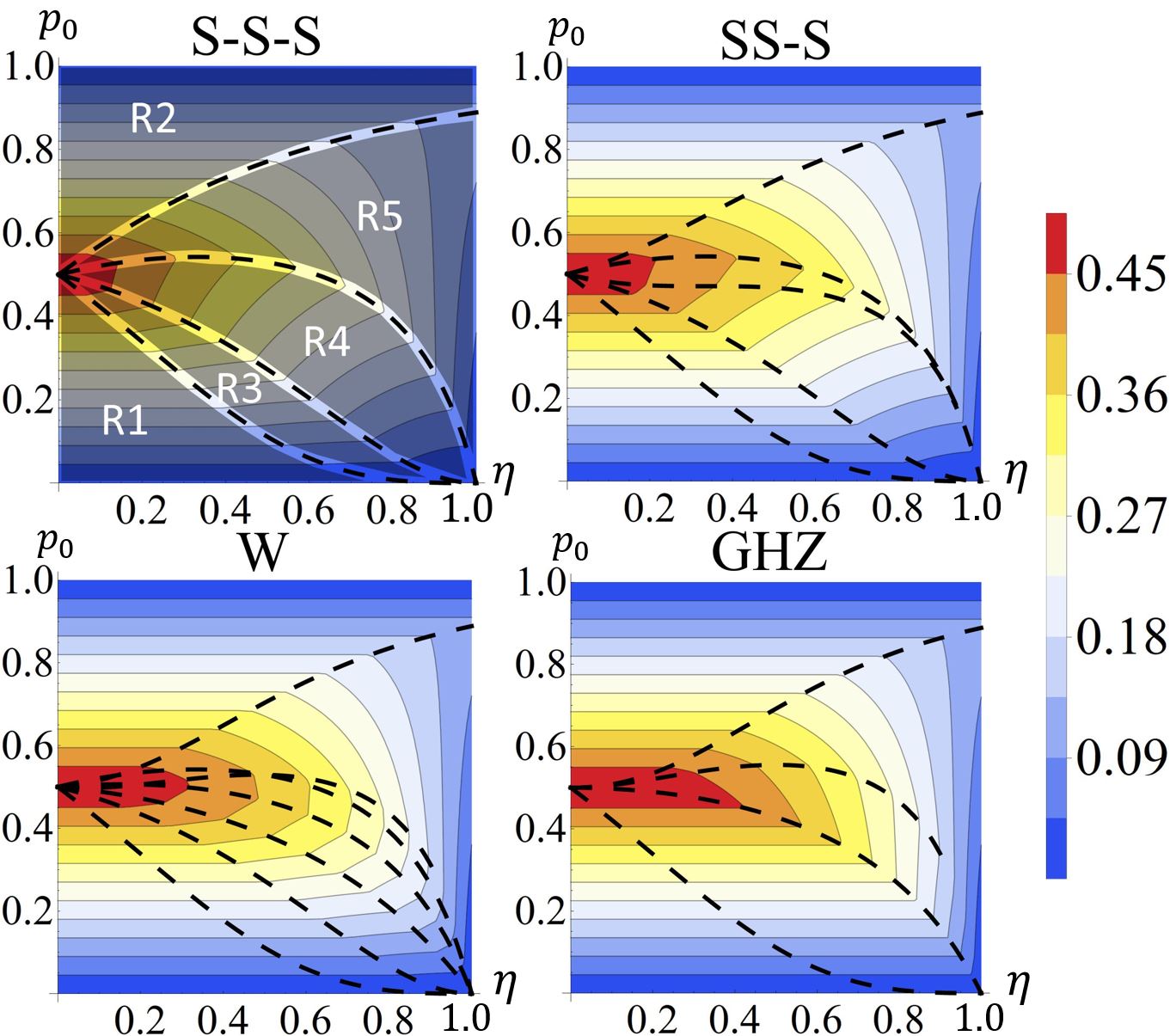}
    \caption{HBs of 3S as a function of $p_0$ and $\eta$.
    The dashed lines represent the boundaries of the analytic solutions for the HBs.}
    \label{3S HB}
\end{figure}

\textbf{Region 5} : $1-\left(\frac{p_0}{p_1}\right)^{\frac{1}{3}}<\eta$ and $1-\frac{1}{\eta^3+3\eta^2+3\eta+2}>p_0\geq1-\frac{1}{-\eta^3-\eta^2+\eta+2}$. The minimal detection error probability is given by 
\begin{equation}
    P_{err} = \frac{1}{8}\left(1+p_1(-6+3\eta+3\eta^2+\eta^3)\right),
\end{equation}
where the optimal POVM is given by
    $\Pi_1 = \ket{\psi}\bra{\psi}$. 
The explicit HBs are drawn in Fig. \ref{3S HB}.


\subsection*{Interpretation of signal-idler entanglement and signal-signal entanglement in the case of two signals and one idler }
\begin{figure}[h!]
    \centering
    \begin{minipage}{0.2\textwidth}
        \centering
        \includegraphics[width=\linewidth]{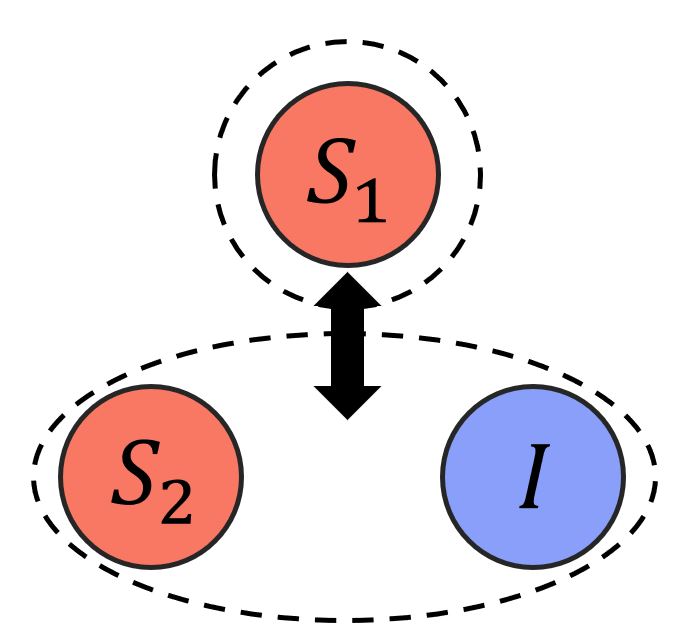}
    \end{minipage}
    \begin{minipage}{0.2\textwidth}
        \centering
        \includegraphics[width=\linewidth]{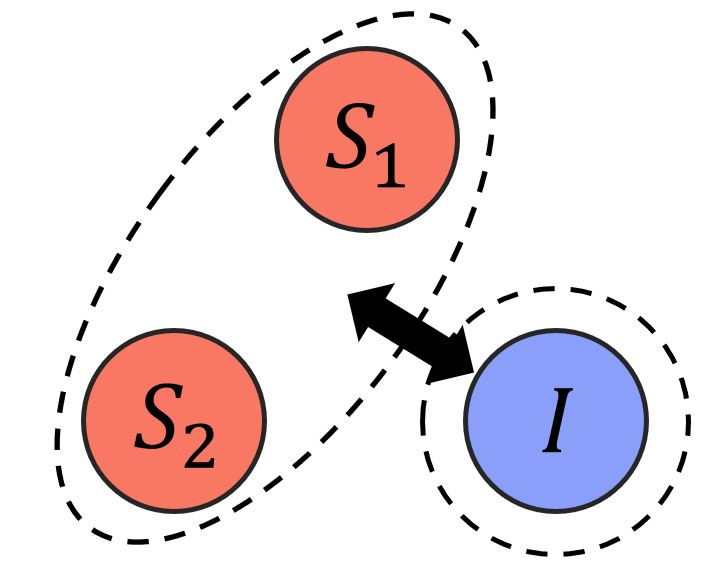}
    \end{minipage}
    \begin{minipage}{0.2\textwidth}
        \centering
        \includegraphics[width=\linewidth]{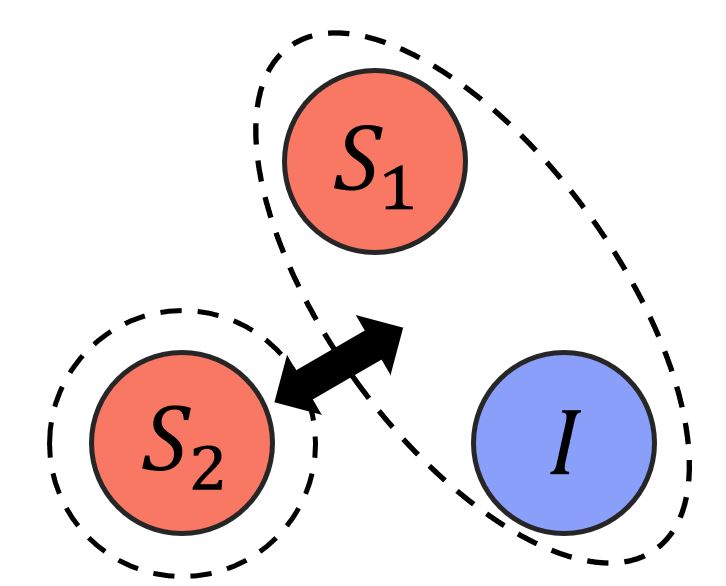}
    \end{minipage}
    \caption{A diagram for three different bipartitions of 2S1I configuration. They represents bipartition $S_1(S_2I)$,$I(S_1S_2)$,$S_2(S_1I)$, respectively. According to our notation, the two qubits enclosed in parentheses form one side of the bipartition, while the remaining qubit constitutes the other side. The arrow between the bipartitions indicates bipartite entanglement.}
    \label{Bipartitions}
\end{figure}

Possible bipartitions are shown in Fig.~\ref{Bipartitions}. Linear entropy $S_L$ is one of the entanglement measure of quantum state, and some authors define it with a normalization \cite{PhysRevA.70.052309,Maleki:19,PhysRevA.61.040101}.
For the states $\ket{\text{S-SI}},\ket{\text{GHZ}},\ket{\text{S-S-I}}$, and $\ket{\text{SS-I}}$, there is a region of having an optimal POVM given as $\Pi_1 = \ket{\psi}\bra{\psi}, \Pi_0 = I-\ket{\psi}\bra{\psi}$. For example, the region 5 of $\ket{\text{S-SI}}$ is such type of region and the corresponding detection error probability is computed as 
\begin{equation}
P_{err} = p_0\bra{\psi}\rho_0\ket{\psi}+p_1(1-\bra{\psi}\rho_1\ket{\psi}).
\end{equation}
Since $\ket{\psi}$ is a three-qubit state, there are three different bipartitions of qubits represented by Schmidt decomposition as : 
    \begin{eqnarray}
        \ket{\psi} = \sum_{k}\sqrt{\alpha_k}\ket{u_k}_{S_1}\otimes\ket{v_k}_{S_2I} 
        = \sum_{m}\sqrt{\beta_m}\ket{u_m}_{S_2}\otimes\ket{v_m}_{S_1I}
        = \sum_{p}\sqrt{\zeta_p}\ket{u_p}_{S_1S_2}\otimes\ket{v_p}_{I}.
        \label{eq:bipartitions}
    \end{eqnarray}
Using the Eq.~(\ref{eq:bipartitions}), we compute two terms $\bra{\psi}\rho_0\ket{\psi}$ and $\bra{\psi}\rho_1\ket{\psi}$ as :
\begin{equation}
    \bra{\psi}\rho_0\ket{\psi} = \frac{1}{4}\left(1-S_L^{I(S_1S_2)}\right),
\end{equation}
    \begin{equation}
        \bra{\psi}\rho_1\ket{\psi} = \eta^2+\frac{\eta(1-\eta)}{2}\left(2-S_L^{S_1(S_2I)}-S_L^{S_2(S_1I)}\right)+\frac{(1-\eta)^2}{4}\left(1-S_L^{I(S_1S_2)}\right),
    \end{equation}
where $S_L^{I(S_1S_2)} = 1-\textrm{Tr}(\textrm{Tr}_{S_1S_2}\left(\ket{\psi}\bra{\psi}\right)^2)$, $S_L^{S_1(S_2I)} = 1- \textrm{Tr}(\textrm{Tr}_{S_1}\left(\ket{\psi}\bra{\psi}\right)^2)$, and $S_L^{S_2(S_1I)} = 1-\textrm{Tr}(\textrm{Tr}_{S_2}\left(\ket{\psi}\bra{\psi}\right)^2)$.
The $S_L^P$ denotes the linear entropy between some bipartition $P$ of three qubits. For example, for the probe state $\ket{\text{S-SI}}$, where the $S_1$ signal qubit is separable, the corresponding bipartite linear entropy is $S_L^{S_1(S_2I)} = 0$, $S_L^{S_2(S_1I)} = 1/2$, and $S_L^{I(S_1S_2)} = 1/2$.

Finally, the detection error probability is computed as :
    \begin{equation}
    P_{err} = p_1(1-\eta) + \frac{\gamma_1}{4}(S_L^{I(S_1S_2)}-1) + \frac{p_1\eta(1-\eta)}{2}(S_L^{S_1(S_2I)}+S_L^{S_2(S_1I)}).
    \label{Error probability using bipartitions}
    \end{equation}

Since $\gamma_1<0$ on illuminable region, we can conclude that entanglement between the bipartition $I(S_1S_2)$ decreases the error probability, whereas entanglement between the bipartitions $S_1(S_2I)$ or $S_2(S_1I)$ increases it. This implies that the performance of QI is worsen by entanglement between signal qubits whereas being enhanced by entanglement between signal and idler qubits. We can explain the order of mean HB between several states in this way, at least in the region where the optimal POVM is given as $\Pi_1 = \ket{\psi}\bra{\psi}$.

In comparison to the performance of a product state, it is physically interpreted as follows. We assume that each signal mode interacts with a target independently. When one signal qubit is lost while the other is reflected, the entanglement between the signal qubits is broken, causing the reflected state to become a mixed state. This effect is quantified by the bipartite linear entropies $S_L^{S_1(S_2I)}$ and $S_L^{S_2(S_1I)}$. Consequently, entanglement between the signal qubits leads to an increase in the miss-detection probability. On the other hand, when both signal qubits are lost, the entanglement between the signal and idler qubits is broken, causing the idler qubit become a mixed state. This effect is quantified by the bipartite linear entropy $S_L^{I(S_1S_2)}$. As a result, entanglement between the signal and idler qubits increases miss-detection probability and decreases the false-alarm probability. However, since $\gamma_1<0$ on illuminable region, the amount of decrements of false-alarm probability is greater than the amount of increments of miss-detection probability.

\subsection*{Holevo information for other configurations}
We compare mean Holevo information with mean HB for QI with three and four qubits \& qudits. The result is provided in Tables~\ref{tab:three qutrit Holevo},~\ref{tab:four qubit Holevo}, and~\ref{tab:four qudit Holevo}. The red boxes indicate that the mean Holevo information is inconsistent with the order of the mean HB. For the QI with three-qutrit, the optimal state is $\text{S-SI}$ state, which has same form with the QI with three-qubit. For the QI with four-qubit, the optimal state is $\text{SI-SI}$ state. For the QI with four-qudit ($d=4$), the optimal state is $\text{SI-SI}$ state. Consider all of the scenarios, the optimal state is $\text{SI-SI}$ state from the QI with four-qudit ($d=4$).

\begin{table}[ht!]
    \centering
    \begin{subtable}{\textwidth}
        \centering
        \scriptsize
        \begin{tabular}{|c|c|c|c|c|c|}
        \hline
        \textbf{2S1I} & \textbf{S-SI} & \textbf{GHZ} & \textbf{S-S-I} & \textbf{SS-I} \\
        \hline
        \textbf{Helstrom} & 0.166427 & 0.179455 & 0.186352 & 0.207634 \\
        \hline
        \textbf{Holevo} & 0.138464 & 0.116811 & 0.10074 & 0.064216 \\
        \hline
        \end{tabular}
        \label{tab:2S1I qutrit Holevo}
    \end{subtable}
    \vspace{1mm}

    \begin{subtable}{\textwidth}
        \centering
        \scriptsize
        \begin{tabular}{|c|c|c|c|c|}
        \hline
        \textbf{3S} & \textbf{S-S-S} & \textbf{S-SS} & \textbf{Cyclic} & \textbf{GHZ} \\
        \hline
        \textbf{Helstrom} & 0.169452 & 0.187244 & 0.203177 & 0.203487 \\
        \hline
        \textbf{Holevo} & 0.134377 & 0.103182 & 0.0760354 & 0.0756214 \\
        \hline
        \end{tabular}
        \label{tab:3S qutrit Holevo}
    \end{subtable}
    \vspace{1mm}

    \begin{subtable}{\textwidth}
        \centering
        \scriptsize
        \begin{tabular}{|c|c|c|c|c|}
        \hline
        \textbf{1S2I} & \textbf{SI-I} & \textbf{GHZ} & \textbf{Cyclic} & \textbf{S-I-I} \\
        \hline
        \textbf{Helstrom} & 0.180758 & 0.180758 & 0.180758 & 0.211189 \\
        \hline
        \textbf{Holevo} & 0.105997 & 0.105997 & 0.105997 & 0.0573067 \\
        \hline
        \end{tabular}
        \label{tab:1S2I qutrit Holevo}
    \end{subtable}
    \caption{Mean HB and mean Holevo information for QI with three-qutrit. Different states are arranged in an increasing order of mean HB as well as a decreasing order of mean Holevo information, from left to right.}
    \label{tab:three qutrit Holevo}
\end{table}

\begin{table}[ht!]
    \begin{subtable}{\textwidth}
        \centering
        \scriptsize
        \begin{tabular}{|c|c|c|c|c|c|c|c|}
        \hline
        \textbf{3S1I} & \textbf{S-S-SI} & \textbf{S-SSI} & \textbf{GHZ} & \textbf{SS-SI} & \textbf{S-S-S-I} & \textbf{S-SS-I} & \textbf{SSS-I} \\
        \hline
        \textbf{Helstrom} & 0.177313 & 0.185091 & 0.18794 & 0.188464 & 0.191024 & 0.204462 & 0.214571\\
        \hline
        \textbf{Holevo} & 0.118623 & 0.105465 & 0.100511 & 0.0986174 & 0.0935365 & 0.0709855 & 0.0532995\\
        \hline
        \end{tabular}
        \label{tab:3S1I qubit Holevo}
    \end{subtable}
    \vspace{1mm}

    \begin{subtable}{\textwidth}
        \centering
        \scriptsize
        \begin{tabular}{|c|c|c|c|c|c|c|}
        \hline
        \textbf{2S2I} & \textbf{SI-SI} & \textbf{S-SI-I} & \textbf{SSI-I} & \textbf{GHZ}  & \textbf{S-S-I-I} & \textbf{SS-I-I} \\
        \hline
        \textbf{Helstrom} & 0.17537 & 0.18816  & 0.19696 & 0.19696  & 0.2058 & 0.22107\\
        \hline
        \textbf{Holevo} & 0.12145 & 0.09682  & 0.0823 & 0.0823  & 0.06745  & 0.04245 \\
        \hline
        \end{tabular}
        \label{tab:2S2I qubit Holevo}
    \end{subtable}
    \vspace{1mm}

    \begin{subtable}{\textwidth}
        \centering
        \scriptsize
        \begin{tabular}{|c|c|c|c|c|c|}
        \hline
        \textbf{4S} & \textbf{S-S-S-S} & \textbf{S-S-SS} & \textbf{SSS-S} & \textbf{SS-SS} & \textbf{GHZ} \\
        \hline
        \textbf{Helstrom} & 0.179423 & 0.191193 & 0.200387 & 0.204333 & 0.206633 \\
        \hline
        \textbf{Holevo} & 0.115666 & 0.0928558 & 0.0822897 & 0.0717958 & 0.070238 \\
        \hline
        \end{tabular}
        \label{tab:4S qubit Holevo}
    \end{subtable}
    \vspace{1mm}

    \begin{subtable}{\textwidth}
        \centering
        \scriptsize
        \begin{tabular}{|c|c|c|c|}
        \hline
        \textbf{1S3I}  & \textbf{SII-I} & \textbf{GHZ}  & \textbf{S-I-I-I} \\
        \hline
        \textbf{Helstrom} & 0.201891 & 0.201891  & 0.225347 \\
        \hline
        \textbf{Holevo}  & 0.0712934 & 0.0712934  & 0.0365761 \\
        \hline
        \end{tabular}
        \label{tab:1S3I qubit Holevo}
    \end{subtable}
    \caption{Mean HB and mean Holevo information for QI with four-qubit. Different states are arranged in an increasing order of mean HB as well as a decreasing order of mean Holevo information, from left to right.}
    \label{tab:four qubit Holevo}
\end{table}

\begin{table}[ht!]
    \begin{subtable}{\textwidth}
        \centering
        \scriptsize
        \begin{tabular}{|c|c|c|c|c|c|c|c|c|}
        \hline
        \textbf{3S1I} & \textbf{S-S-SI} & \textbf{S-SSI} & \textbf{SS-SI} & \textbf{GHZ} & \textbf{S-S-S-I} & \textbf{Cyclic} & \textbf{S-SS-I} & \textbf{SSS-I} \\
        \hline
        \textbf{Helstrom} & 0.14128 & 0.15310 & \cellcolor{red!30}0.15559 & \cellcolor{red!30}0.15666 & 0.15711 & 0.16500 & 0.17753 & 0.19828\\
        \hline
        \textbf{Holevo} & 0.19014 & 0.16892 & \cellcolor{red!30}0.16081 & \cellcolor{red!30}0.16215 & 0.15852 & 0.14686 & 0.12198 & 0.08794 \\
        \hline
        \end{tabular}
        \label{tab:3S1I qu4it Holevo}
    \end{subtable}
    \vspace{1mm}

    \begin{subtable}{\textwidth}
        \centering
        \tiny
        \begin{tabular}{|c|c|c|c|c|c|c|c|}
        \hline
        \textbf{2S2I} & \textbf{SI-SI} & \textbf{S-SI-I}  & \textbf{Cyclic} & \textbf{SSI-I} & \textbf{GHZ}  & \textbf{S-S-I-I}  & \textbf{SS-I-I} \\
        \hline
        \textbf{Helstrom} & 0.13803 & 0.15518  & 0.16125 & 0.17069 & 0.17069  & 0.17537  & 0.2009 \\
        \hline
        \textbf{Holevo} & 0.19357 & 0.16158  & 0.15304 & 0.13592 & 0.13592  & 0.12145  & 0.07694 \\
        \hline
        \end{tabular}
        \label{tab:2S2I qu4it Holevo}
    \end{subtable}
    \vspace{1mm}

    \begin{subtable}{\textwidth}
        \centering
        \scriptsize
        \begin{tabular}{|c|c|c|c|c|c|}
        \hline
        \textbf{4S} & \textbf{S-S-S-S} & \textbf{S-S-SS} & \textbf{SS-SS} & \textbf{SSS-S} & \textbf{GHZ} \\
        \hline
        \textbf{Helstrom} & 0.142971 & 0.158986 & 0.174428 & 0.178212 & 0.185839 \\
        \hline
        \textbf{Holevo} & 0.185751 & 0.151839 & 0.134266 & 0.117413 & 0.113169 \\
        \hline
        \end{tabular}
        \label{tab:4S qu4it Holevo}
    \end{subtable}
    \vspace{1mm}

    \begin{subtable}{\textwidth}
        \centering
        \scriptsize
        \begin{tabular}{|c|c|c|c|}
        \hline
        \textbf{1S3I}  & \textbf{SII-I} & \textbf{GHZ}  & \textbf{S-I-I-I} \\
        \hline
        \textbf{Helstrom}  & 0.170629 & 0.170629  & 0.201891 \\
        \hline
        \textbf{Holevo}  & 0.125588 & 0.125588  & 0.0712934 \\
        \hline
        \end{tabular}
        \label{tab:1S3I qu4it Holevo}
    \end{subtable}
    \caption{Mean HB and mean Holevo information for QI with four-qudit (d=4). Different states are arranged in an increasing order of mean HB as well as a decreasing order of mean Holevo information, from left to right.}
    \label{tab:four qudit Holevo}
\end{table}

\section*{Acknowledgement}
This work was supported by the Defense Acquisition Program Administration and the Agency for Defense Development.

\section*{Author contributions statement}

S.-Y. L. and S.K. initiated the project. S.K. conducted the simulations. All authors analysed the results and reviewed the manuscript.


\begin{thebibliography}{99}

\bibitem{qubit} Schumacher, B. Quantum coding. \textit{Phys. Rev. A} \textbf{51}, 2738-2747 (1995).

\bibitem{EPR} Einstein, A., Podolsky, B. \& Rosen, N. Can Quantum-Mechanical Description of Physical Reality Be Considered Complete? \textit{Phys. Rev.} \textbf{47}, 770-780 (1935).

\bibitem{Lloyd08} Lloyd, S. Enhanced Sensitivity of Photodetection via Quantum Illumination. \textit{Science} \textbf{321}, 1463-1465 (2008).

\bibitem{Audenaert07} Audenaert, K. M. R. \textit{et al.} Discriminating States: The Quantum Chernoff Bound. \textit{Phys. Rev. Lett.} \textbf{98}, 1463-1465 (2007).

\bibitem{Calsa08} Calsamiglia, J., Mu\~noz-Tapia, R., Masanes, Ll., Acin, A. \& Bagan, E. Quantum Chernoff bound as a measure of distinguishability between density matrices: Application to qubit and Gaussian states. \textit{Phys. Rev. A} \textbf{77}, 032311 (2008).

\bibitem{Stefano08} Pirandola, S. \& Lloyd, S. Computable bounds for the discrimination of Gaussian states. \textit{Phys. Rev. A} \textbf{78}, 012331 (2008).

\bibitem{SL09} Shapiro, J. H. \& Lloyd, S. Quantum illumination versus coherent-state target detection. \textit{New J. Phys.} \textbf{11}, 063045 (2009).

\bibitem{Tan08} Tan, S.-H. \textit{et al.} Quantum illumination with Gaussian states. \textit{Phys. Rev. Lett.} \textbf{101}, 253601 (2008).

\bibitem{Sanz} Sanz, M., Las Heras, U., Garc\'{\i}a-Ripoll, J. J., Solano, E. \& Di Candia, R. Quantum Estimation Methods for Quantum Illumination. \textit{Phys. Rev. Lett.} \textbf{118}, 070803 (2017).

\bibitem{Karsa} Karsa, A., Spedalieri, G., Zhuang, Q. \& Pirandola, S. Quantum illumination with a generic Gaussian source. \textit{Phys.Rev.Res.} \textbf{2}, 023414 (2020).

\bibitem{Lee} Lee, S.-Y., Ihn, Y.S. \& Kim, Z. Quantum illumination via quantum-enhanced sensing. \textit{Phys. Rev. A} \textbf{103}, 012411 (2021).

\bibitem{Prabhu21} Prabhu, A. V., Suri, B. \& Chandrashekar, C. M. Hyperentanglement-enhanced quantum illumination. \textit{Phys. Rev. A} \textbf{103}, 052608 (2021).

\bibitem{Noh} Noh, C., Lee, C. \& Lee, S.-Y. Quantum illumination with definite photon-number entangled states. \textit{J. Opt. Soc. Am. B} \textbf{39}, 1316 (2022).

\bibitem{Lee24} Lee, S.-Y., Kim, J., Kim, Z. \& Kim, D.Y. Asymmetric entanglement for quantum target sensing. \textit{Phys. Rev. A} \textbf{109}, 042429 (2024).

\bibitem{Pannu} Pannu, A., Helmy, A.S. \& Gamal, H. El. Quantum illumination with high-dimensional Bell states. \textit{Phys. Rev. A} \textbf{110}, L050603 (2024).

\bibitem{Jung} Jung, E. \& Park, D. Quantum illumination with three-mode Gaussian state. \textit{Quantum Inf. Process.} \textbf{21}, 71 (2021).

\bibitem{Gallego} Torrom\'{e}, R. G. Quantum Illumination with Multiple Entangled Photons. \textit{Adv. Quantum Technol.} \textbf{4}, 2100101 (2021).

\bibitem{DePalma18} De Palma, G. \& Borregaard, J. Minimum error probability of quantum illumination. \textit{Phys. Rev. A} \textbf{98}, 012101 (2018).

\bibitem{Pirandola} Pirandola, S., Laurenza, R., Lupo, C. \& Pereira, J.L. Fundamental limits to quantum channel discrimination. \textit{npj Quantum Inform.} \textbf{5}, 50 (2019).

\bibitem{Calsa10} Calsamiglia, J., de Vicente, J.I., Mu\~noz-Tapia, R. \& Bagan, E. Local Discrimination of Mixed States. \textit{Phys. Rev. Lett.} \textbf{105}, 080504 (2010).

\bibitem{Bae15} Bae, J. \& Kwek, L.-C. Quantum state discrimination and its applications. \textit{J. Phys. A-Math. Gen.} \textbf{48}, 083001 (2015).

\bibitem{Yung20} Yung, M.-H., Meng, F., Zhang, X.-M. \& Zhao, M.-J. One-shot detection limits of quantum illumination with discrete signals. \textit{npj Quantum Inform.} \textbf{6}, 75 (2020).

\bibitem{Xu21} Xu, F. \textit{et al.} Experimental Quantum Target Detection Approaching the Fundamental Helstrom Limit. \textit{Phys. Rev. Lett.} \textbf{127}, 040504 (2021).

\bibitem{GHSZ} Greenberger, D. M., Horne, M.A., Shimony, A. \& Zeilinger, A. Bell's theorem without inequalities. \textit{Am. J. Phys.} \textbf{58}, 1131-1143 (1990).

\bibitem{DVC00} D\"{u}r, W., Vidal, G. \& Cirac, J.I. Three qubits can be entangled in two inequivalent ways. \textit{Phys. Rev. A} \textbf{62}, 062314 (2000).

\bibitem{Aguilar19} Aguilar, G. H. \textit{et al.} Experimental investigation of linear-optics-based quantum target detection. \textit{Phys. Rev. A} \textbf{99}, 053813 (2019).

\bibitem{Weedbrook} Weedbrook, C., Pirandola, S., Thompson, J., Vedral, V. \& Gu, M. How discord underlies the noise resilience of quantum illumination. \textit{New J. Phys.} \textbf{18}, 043027 (2016).

\bibitem{Helstrom} Helstrom, C. W. Quantum detection and estimation theory. \textit{J. Stat. Phys.} \textbf{1}, 231-252 (1969).

\bibitem{Cabello} Cabello, A. $N$-Particle $N$-Level Singlet States: Some Properties and Applications. \textit{Phys. Rev. Lett.} \textbf{89}, 100402 (2002).

\bibitem{Jo} Jo, Y., Bae, K. \& Son W. Enhanced Bell state measurement for efficient measurement-device-independent quantum key distribution using 3-dimensional quantum states. \textit{Sci. Rep.} \textbf{9}, 687 (2019).

\bibitem{Jo25} Jo, Y. \textit{et al.} Security analysis of qutrit quantum secret sharing with linear optical correlation measurement. \textit{Sci. Rep.} \textbf{15}, 19836 (2025).

\bibitem{Pan} Luo, Y.-H. \textit{et al.} Quantum Teleportation in High Dimensions. \textit{Phys. Rev. Lett.} \textbf{123}, 070505 (2019).

\bibitem{Sai} Subramanian, M. \& Vinjanampathy, S. Shallow-depth variational quantum hypothesis testing. \textit{Phys. Rev. A} \textbf{110}, 032424 (2024).

\bibitem{PhysRevA.70.052309} Peters, N. A., Wei, T.-C. \& Kwiat, P. G. Mixed-state sensitivity of several quantum-information benchmarks. \textit{Phys. Rev. A} \textbf{70}, 052309 (2004).

\bibitem{Maleki:19} Maleki, Y. \& Zheltikov, A. M. Linear entropy of multiqutrit nonorthogonal states. \textit{Opt. Express} \textbf{27}, 8291-8307 (2019).

\bibitem{PhysRevA.61.040101} Bose, S. \& Vedral, V. Mixedness and teleportation. \textit{Phys. Rev. A} \textbf{61}, 040101 (2000).

\end{thebibliography}
\end{document}